\documentclass[twocolumn,showpacs,prb,superscriptaddress, amsmath, amssymb, 10pt]{revtex4-2}

\usepackage{mathtools}        
\usepackage{bbm}
\usepackage{float}            
\usepackage{graphicx}         
\usepackage[dvipsnames]{xcolor}
\usepackage[caption=false]{subfig}

\usepackage{physics}
\usepackage{relsize}          
\usepackage{epstopdf}
\usepackage{soul}
\usepackage{hyperref}
\hypersetup{
	linkcolor=blue,
	citecolor=blue,
	filecolor=blue,
	urlcolor=blue,
	colorlinks=true
}

\usepackage{hhline}
\usepackage{lipsum}

\usepackage{pifont}      
\usepackage{color}       
\usepackage{framed}		 
\definecolor{shadecolor}{gray}{0.965}

\renewcommand{\v}[1]{\ensuremath{\boldsymbol{\mathbf{#1}}}}

\newcommand{\UIUCPHYS}[0]{Department of Physics, University of Illinois at Urbana-Champaign, Urbana, IL 61801, USA}
\newcommand{\UICPHYS}[0]{Department of Physics, University of Illinois at Chicago, Chicago, IL 60607, USA}

\newcommand*{\figref}[2][]{%
  \hyperlink{fig:#2}{%
    \ref*{fig:#2}%
    \ifx \\#1\\%
    \else
      \textbf{#1}%
    \fi
  }%
}

\newcommand*{\subfigref}[2][]{%
  \hyperlink{fig:#2}{%
    \ifx \\#1\\%
    \else
      \textbf{#1}%
    \fi
  }%
}
\DeclareCaptionLabelFormat{noparens}{\textbf{#2}}
\allowdisplaybreaks 

\begin{document}

\title{Higher Order Topological Superconductivity in Magnet-Superconductor Hybrid Systems}

\author{Ka Ho Wong}\affiliation{\UICPHYS}
\author{Mark R. Hirsbrunner}\affiliation{\UIUCPHYS}
\author{Jacopo Gliozzi}\affiliation{\UIUCPHYS}
\author{Arbaz Malik}\affiliation{\UICPHYS}
\author{Barry Bradlyn}\affiliation{\UIUCPHYS}
\author{Taylor L. Hughes}\affiliation{\UIUCPHYS}
\author{Dirk K. Morr}\affiliation{\UICPHYS}

\begin{abstract}
Quantum engineering of topological superconductors and of the ensuing Majorana zero modes might hold the key for realizing a new paradigm for the implementation of topological quantum computing and topology-based devices. Magnet-superconductor hybrid (MSH) systems have proven to be experimentally versatile platforms for the creation of topological superconductivity by custom-designing the complex structure of their magnetic layer. Here, we demonstrate that higher order topological superconductivity (HOTSC) can be realized in two-dimensional MSH systems by using stacked magnetic structures. We show that the sensitivity of the HOTSC to the particular magnetic stacking opens an unprecedented ability to tune the system between trivial and topological phases using atomic manipulation techniques. We propose that the realization of HOTSC in MSH systems, and in particular the existence of the characteristic Majorana corner modes, allows for the implementation of a measurement-based protocols for topological quantum computing.
\end{abstract}

\maketitle

\noindent {\bf Introduction}\\
Topological superconductors present a new paradigm for the realization of quantum computing as they harbor Majorana zero modes (MZMs), whose non-Abelian statistics and robustness against decoherence and disorder are well suited for the implementation of quantum gates~\cite{Nayak2008}. Magnet superconductor hybrid (MSH) systems, in which magnetic adatoms are placed on the surface of $s$-wave superconductors, have proven to be a versatile platform for the creation of topological superconductors. Strong evidence for the existence of MZMs has been reported in ferromagnetic one-dimensional (1D)~\cite{Nadj-Perge2014,Ruby2015,Pawlak2016,Kim2018} and two-dimensional (2D)~\cite{Menard2017,Palacio-Morales2019,Kezilebieke2020} MSH structures. Moreover, the capability to quantum engineer more complex magnetic structure in MSH systems, ranging from skyrmionic~\cite{Mascot2021} to antiferromagnetic and 3{\bf Q}-structures~\cite{Bedow2020}, is predicted to provide new venues for creating, tuning and manipulating topological phases. Indeed, antiferromagnetic MSH systems were recently shown to give rise to nodal topological superconductivity~\cite{Bazarnik2022}. The observation of strong topological superconducting phases in 2D MSH systems raises the question: can higher order topological superconductor (HOTSC) phases, in which MZMs emerge as localized corner modes~\cite{wang2019,khalaf2018higher,ono2020refined,ono2021z,schindler2020pairing,tang2022high,langbehn2017reflection,hsu2018majorana}, be realized with more complex magnetic structures?

In this article, we answer this question in the affirmative. We demonstrate that HOTSCs, in addition to strong and weak topological superconducting (TSC) phases, can be quantum engineered in 2D MSH systems using a spatially-modulated magnetic structure. We find that the resulting HOTSC phase possesses extrinsic higher order topology and emerges as a boundary-obstructed phase \cite{benalcazar2017science,benalcazar2017prb,Geier2018,Khalaf2021}. Such phases are separated from trivial phases by surface, rather than bulk, gap closings. The MZMs that appear in this phase are geometrically determined by the shape and termination of the magnetic structure. As a result, the HOTSC phase provides an unprecedented opportunity to use atomic manipulation techniques to generate and fuse MZMs, and to tune between trivial and topological phases. To characterize the phases of our MSH system we identify key spectroscopic signatures of all three topological phases -- with the HOTSC phase exhibiting distinctive Majorana corner modes -- that can be observed using scanning tunneling spectroscopy. Furthermore, we identify the microscopic mechanisms responsible for the emergence of HOTSC and weak TSC phases, and show that these phases are generic features of stacked magnetic structures. Finally, we propose that the unique structure of Majorana corner modes in the HOTSC phase provides a new venue for the realization of topological quantum gates. Our results thus provide a blueprint for the quantum engineering of an HOTSC phase in MSH structures.\\

\noindent  {\bf Results}\\
{\bf Theoretical Model}

To investigate topological superconducting phases in two-dimensional MSH structures, we consider arrays of magnetic adatoms placed on the surface of a conventional $s$-wave superconductor, as described by the  Hamiltonian~\cite{Ron2015,Li2016,Rachel2017}
\begin{equation}
    \begin{aligned}
        \mathcal{H} =& -t \sum_{\v{r}, \v{\delta},  \alpha} c^\dagger_{\v{r}, \alpha} c_{\v{r}+\v{\delta}, \alpha} - \mu \sum_{\v{r}, \alpha} c^\dagger_{\v{r}, \alpha} c_{\v{r}, \alpha} \\
        &+ i \lambda \sum_{\substack{ \v{r}, \v{\delta} \\ \alpha, \beta}} c^\dagger_{\v{r}, \alpha} \left(  \left[\v{\delta} \times \v{\sigma} \right] \cdot \hat{z} \right)_{\alpha, \beta}  c_{\v{r} + \v{\delta}, \beta} \\
        &+ \Delta \sum_{\v{r}} \left( c^\dagger_{\v{r}, \uparrow} c^\dagger_{\v{r}, \downarrow} + c_{\v{r}, \downarrow} c_{\v{r}, \uparrow} \right) \\
        &+ J {\sum_{\substack{\v{R} \\ \alpha, \beta}}}^\prime c^\dagger_{\v{R}, \alpha} \left[ \v{S}_{\bf R} \cdot \v{\sigma} \right]_{\alpha,\beta} c_{\v{R}, \beta} \; .
    \label{eq:H}
    \end{aligned}
\end{equation}
\begin{figure*}[htb]
    \centering
    \subfloat[]{\label{fig:1a}\includegraphics[height=2in]{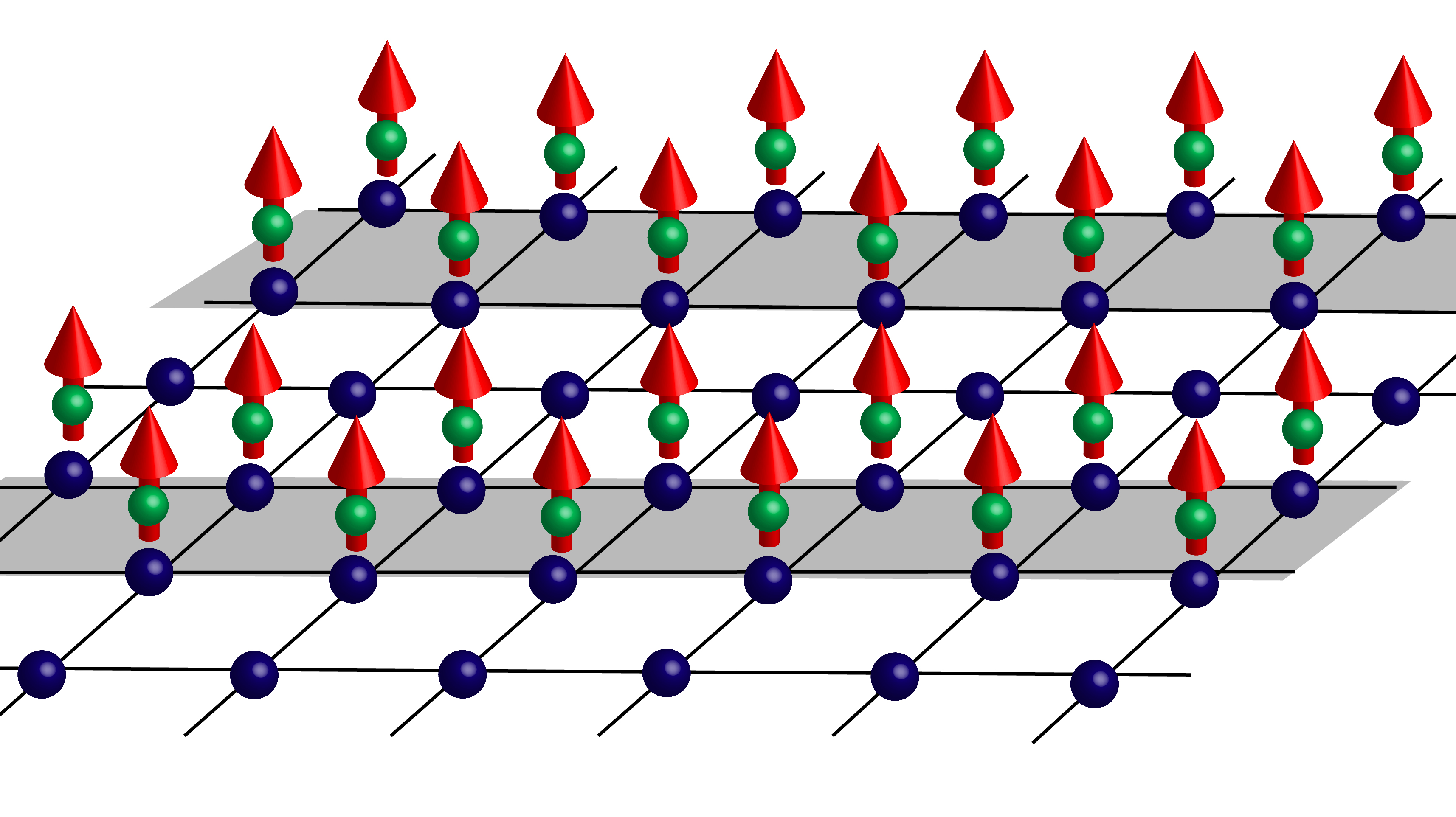}}
    \hspace{0.2in}
    \subfloat[]{\label{fig:1b}\includegraphics[height=2in]{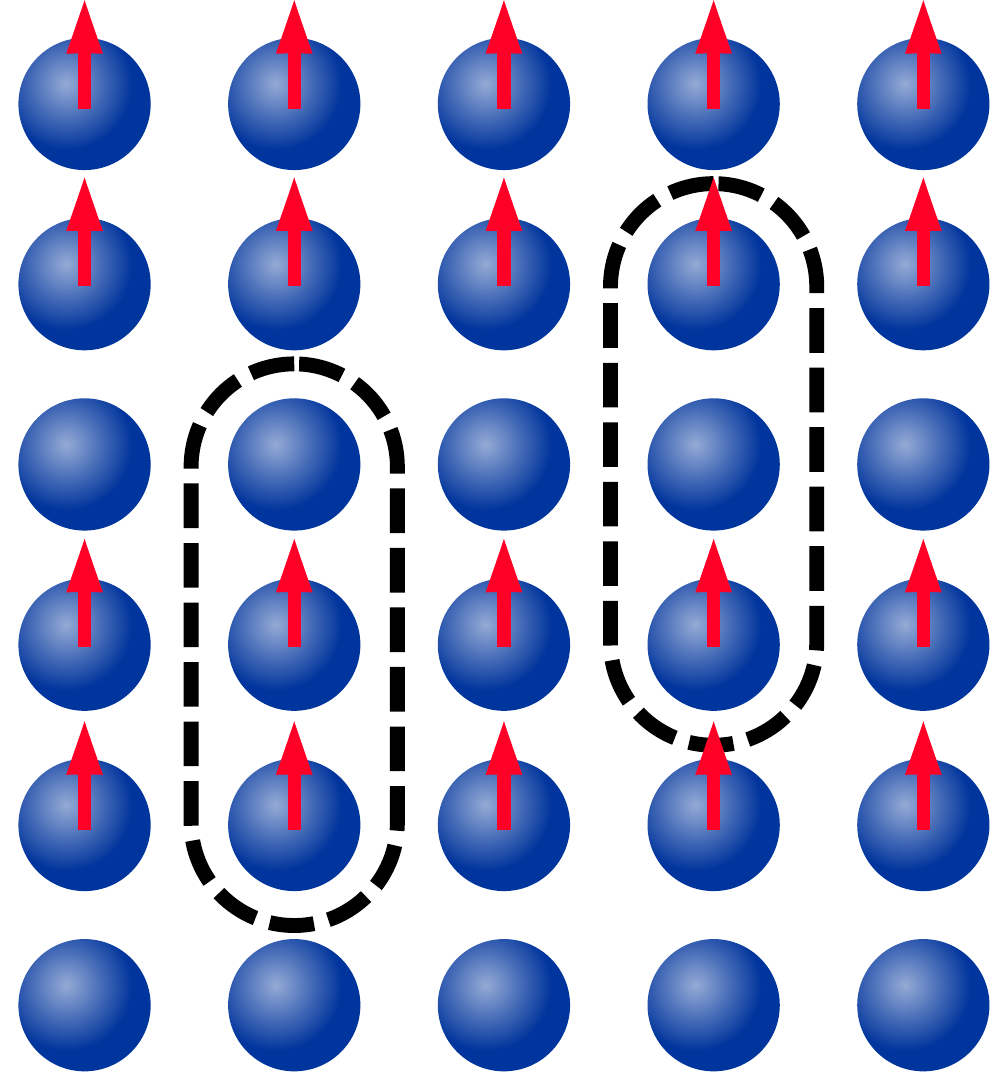}}
    \\
    \subfloat[]{\label{fig:1c}\includegraphics[width=0.4\textwidth]{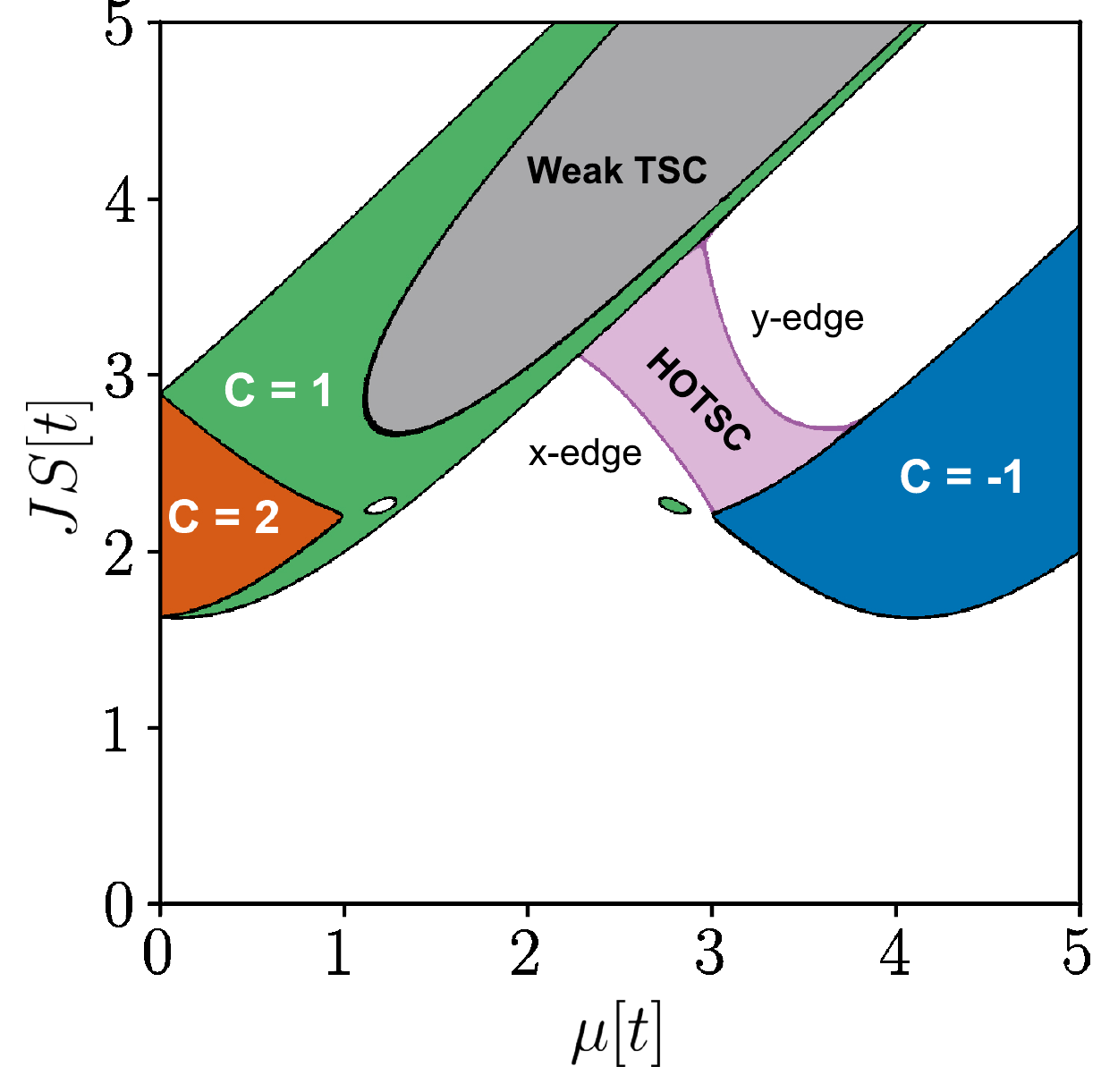}}
    \hspace{0.05\textwidth}
    \subfloat[]{\label{fig:1d}\includegraphics[width=0.4\textwidth]{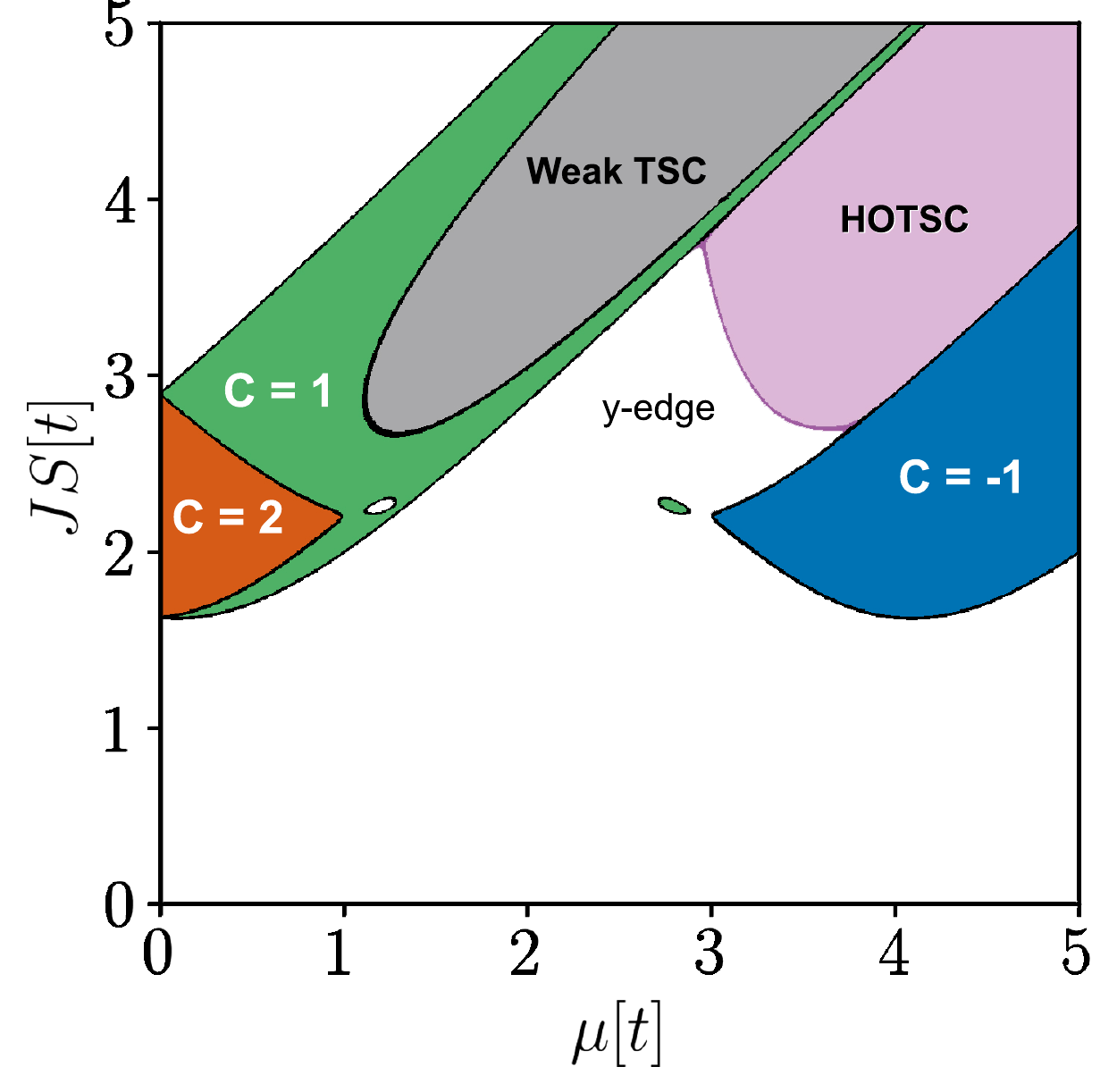}}
    \caption[]{{\bf MSH systems with stacked magnetic structures} {\bf a} Schematic illustration of the stacked MSH system with green (blue) spheres denoting the magnetic adatoms (the superconducting sites), and red arrows indicating the adatom spins. The gray area shows the superconducting region covered by magnetic adatoms.  {\bf b} A top-down view of the MSH system. Filled blue circles represent lattice sites of the superconducting substrate and red arrows indicate the magnetic adatoms. The dashed lines indicate two unit  choices that represent the stackings $(J, J, 0)$ (left) and $(J, 0, J)$ (right). Topological phase diagram of an MSH system with {\bf c}  $(J, 0, J)$ stacking and {\bf d} $(J, J, 0)$ stacking as a function of $\mu$ and $JS$. The bulk and edge gap closings are denoted by black and purple lines, respectively. Parameters are $\lambda=0.8t$ and $\Delta=1.2t$. The system size $(N_i)$ and the island size $(W_i)$ used for calculating the $x$-edge and $y$-edge gap closings are $(N_y,W_y)=(120,80)$ and $(N_x,W_x)=(280,240)$ respectively.}
    \label{fig:1}
\end{figure*}
Here,  $c^\dagger_{\v{r}, \alpha}$ creates an electron with spin $\alpha$
at site $\v{r}$, $-t$ is the nearest-neighbor hopping amplitude on a 2D square lattice, $\mu$ is the chemical potential, $\lambda$ is the Rashba spin-orbit coupling strength, $\Delta$ is the $s$-wave superconducting order parameter, $\v{\sigma}$ is the vector of Pauli-spin matrices, and the vectors $\v{\delta}$ connect nearest-neighbor sites. Moreover, $J$ is the amplitude of the magnetic exchange coupling between the adatom spin $\v{S}_{\v{R}}$ (with magnitude $S$) located at site $\v{R}$ and the conduction electrons, where the primed sum runs over all sites $\v{R}$ decorated with magnetic adatoms. As Kondo screening is suppressed due to the hard superconducting gap~\cite{Balatsky2006,Heinrich2018}, we consider the magnetic adatoms to be classical in nature with an out-of-plane ferromagnetic alignment.
To simplify the discussion of the topological phase diagram and to identify spatially well-localized MZMs, we consider below a somewhat large superconducting order parameter of $\Delta = 1.2 t$. However, the qualitative features of the phase diagram persist to smaller, more realistic values of $\Delta$, as shown in Supplementary Note 1.

While superconducting surfaces uniformly covered by magnetic adatoms have been considered before~\cite{Ron2015,Li2016,Rachel2017}, here we consider MSH structures with spatially-modulated adatom configurations. These configurations are comprised of adatom chains parallel to the $x$-axis that are arrayed in the $y$-direction, as schematically shown in Fig.~\ref{fig:1a}. In particular, we consider a system in which two adjacent chains of magnetic adatoms are followed by an empty row of the superconducting substrate (i.e., a chain not covered by magnetic adatoms). As illustrated in Fig.~\ref{fig:1b}, finite islands of such MSH systems can be terminated in the $y$-direction by a single or double row of magnetic adatoms, which we refer to as a $(J,0,J)$ or $(J,J,0)$ stacking, respectively.

\begin{figure*}[!ht]
    \centering
    \hypertarget{fig:2}{}
    \includegraphics[width=0.7\textwidth]{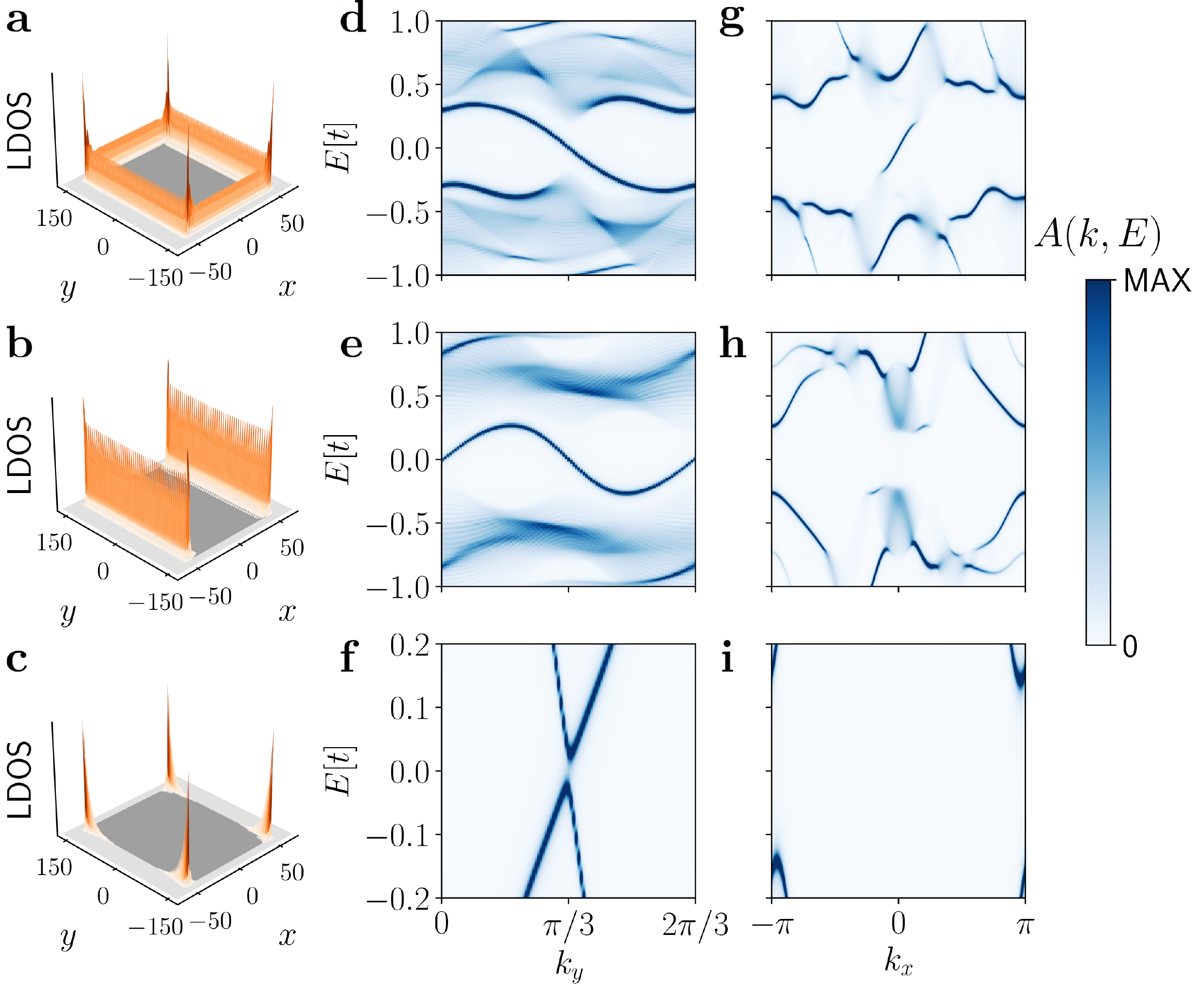}
    \caption{{\bf Spectroscopic signatures of topological phases}  Zero-energy LDOS for a finite magnetic island (left column), $y$-edge spectral function (center column), and $x$-edge spectral function (right column) of the strong TSC phase (top row), weak TSC phase (middle row), and HOTSC phase (bottom row). All calculations were performed for an MSH system with a $(J, 0, J)$ stacking. Parameters are $(\mu,\lambda,\Delta,JS)=(0.5, 0.8, 1.2, 3)t$ for the top row, $(\mu,\lambda,\Delta,JS)=(2, 0.8, 1.2, 4)t$ for the middle row, and $(\mu,\lambda,\Delta,JS)=(2.8,0.8,1.2,3)t$ for the bottom row.}
    \label{fig:2}
\end{figure*}

Since the Hamiltonian in Eq.~(\ref{eq:H}) breaks time reversal symmetry, the MSH system belongs to the Altland-Zirnbauer class D~\cite{altlandzirnbauer, kitaev2009periodic, ryu2010topological}. For a two-dimensional system, the corresponding topological invariant is the Chern number,
which can be computed via~\cite{kitaev2009periodic, ryu2010topological, avron1983homotopy}
\begin{align}\label{eq:C}
 C & =  \frac{1}{2\pi i} \int_{\text{BZ}} d^2k \mathrm{Tr} ( P_{\bf{k}} [ \partial_{k_x} P_{\bf{k}}, \partial_{k_y} P_{\bf{k}} ] )  \nonumber \\
 P_{\bf{k}} & = \sum_{E_n(\bf{k}) < 0} |\Psi_n({\bf{k}}) \rangle \langle \Psi_n({\bf{k}})|,
\end{align}
where $E_n({\bf{k}})$ and $|\Psi_n({\bf{k}}) \rangle$ are the eigenenergies and the eigenvectors of the Hamiltonian in Eq.~(\ref{eq:H}). The index $n$ and the trace both run over the spin, magnetic sublattice, and Nambu degrees of freedom.

In addition to the Chern number, lattice translation symmetry permits the definition of two $\mathbb{Z}_2$-valued weak invariants for 2D superconductors in symmetry class D, which we call $v_{x,\pi}$, and $v_{y,\pi}$~\cite{seroussi2014topological, fu2007topological}.
In contrast to the Chern number, which is associated with an isotropic topological response, weak invariants indicate the presence of gapless surface states on only certain edges. In particular, when $C=0$, non-trivial values of $v_{x,\pi} (v_{y,\pi})$ indicate edge modes on edges parallel to the $x$-direction ($y$-direction). We show below that in certain parts of the phase diagram where the Chern number vanishes, one of the weak invariants is non-zero, reflecting the existence of a weak topological phase.\\

\noindent {\bf Topological Phase Diagram}\\
In Fig.~\ref{fig:1c} we present the topological phase diagram of the $(J,0,J)$-stacked MSH structure in the $(\mu,JS)$-plane.
In addition to a trivial phase and three strong topological phases with non-zero Chern numbers $C$,
we also identify a region in the phase diagram having $C=0$ and $v_{y,\pi} \not = 0$, i.e., a weak topological superconductor~\cite{seroussi2014topological, fu2007topological}. The transitions between the trivial, weak, and strong phases are associated with bulk gap closings (see the black lines in Fig.~\ref{fig:1c} and ~\ref{fig:1d}), but interestingly, when we consider the phase diagram for MSH structures with a finite spatial extent in either the $x$- or $y$-directions (exhibiting edges we refer to as $y$-edges or $x$-edges, respectively) new gap closings appear on the edges. For the $(J,0,J)$ stacking, the MSH system of Fig.~\ref{fig:1a} exhibits edge gap closings (denoted by purple lines in Fig.~\ref{fig:1c}) on both the $x$- and $y$-edges. Previous work has shown that such edge transitions can give rise to boundary obstructed topological phases having extrinsic higher order topology~\cite{benalcazar2017science,benalcazar2017prb,Geier2018,Khalaf2021}. Below we demonstrate that the phase bounded by these edge gap closings is an extrinsic, higher order topological superconductor (HOTSC). Furthermore, we find that changing the $(J,0,J)$ stacking to a $(J,J,0)$ stacking eliminates the $x$-edge gap closing and shifts the phase we identify as the HOTSC, as shown in Fig.~\ref{fig:1d}. This sensitivity of the gap closing transitions to the MSH edge termination is consistent with the existence of an extrinsic boundary obstructed phase~\cite{Geier2018,Khalaf2021}.\\

\begin{figure*}[!htb]
    \captionsetup[subfigure]{captionskip=-5pt, position=top, font={bf,Large},singlelinecheck=off,labelformat=noparens,justification=raggedright}
    \centering
    \subfloat[][]{\label{fig:3a}\includegraphics[width=0.33\textwidth]{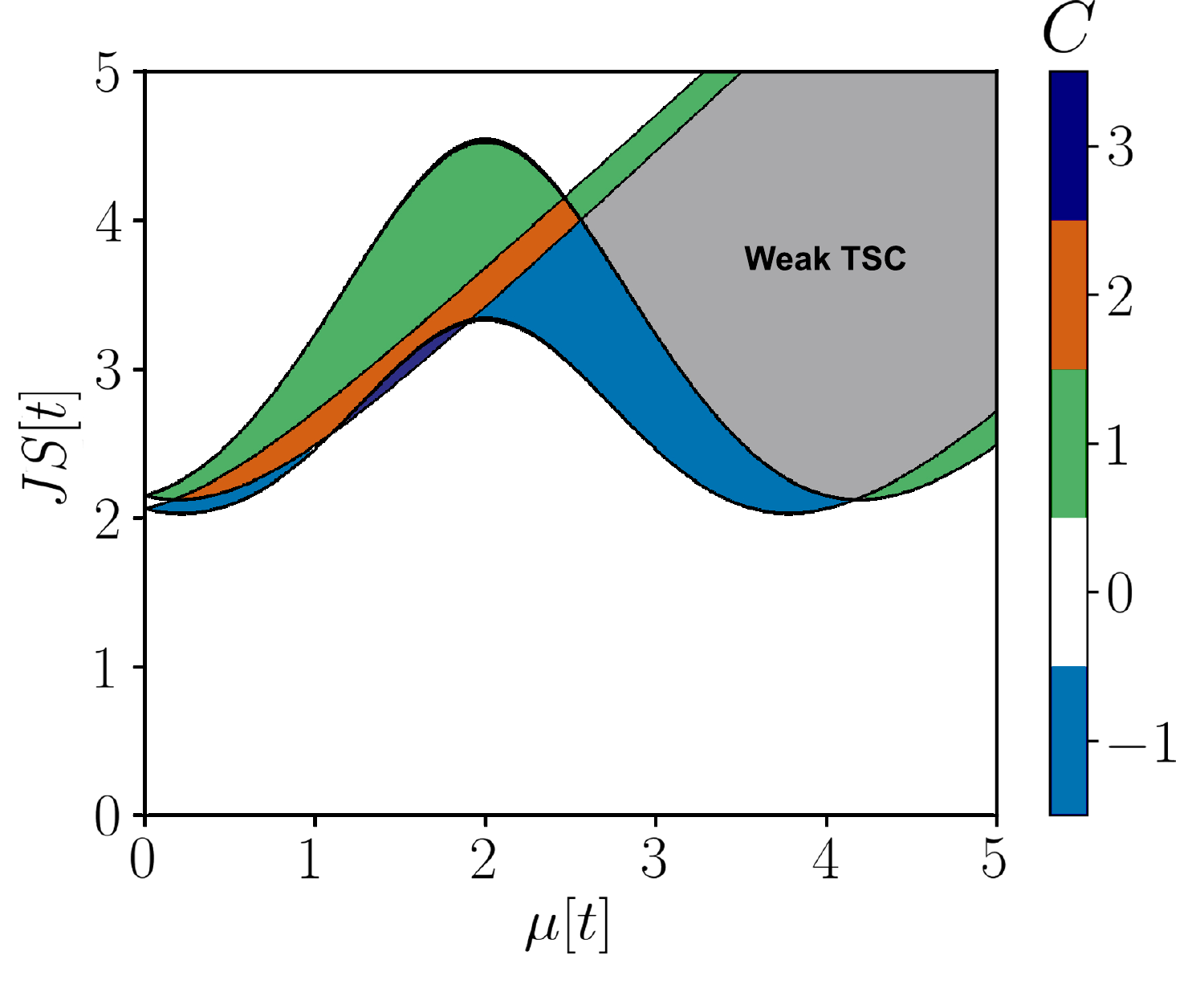}}
    \subfloat[][]{\label{fig:3b}\includegraphics[width=0.33\textwidth]{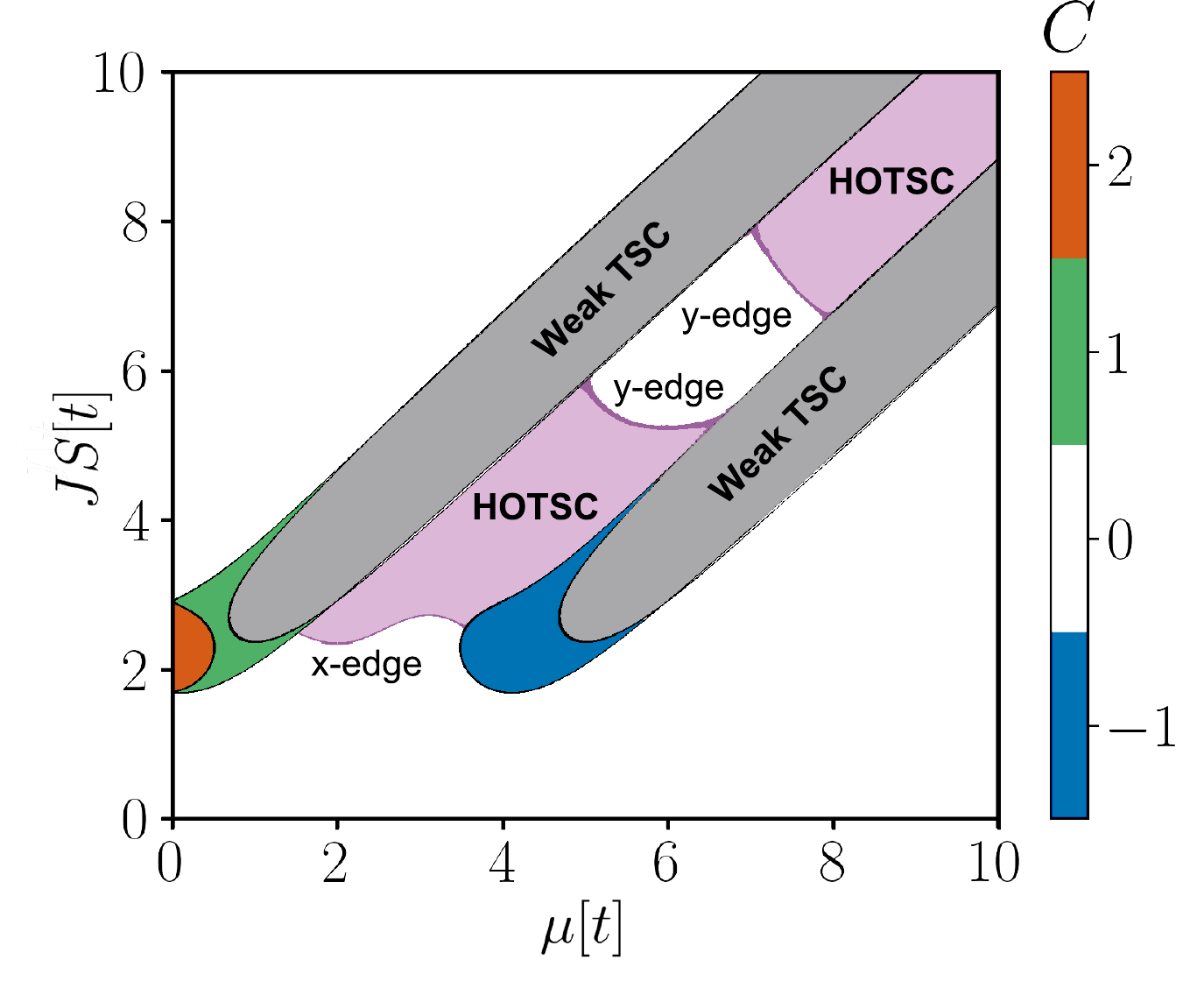}}
    \subfloat[][]{\label{fig:3c}\includegraphics[width=0.33\textwidth]{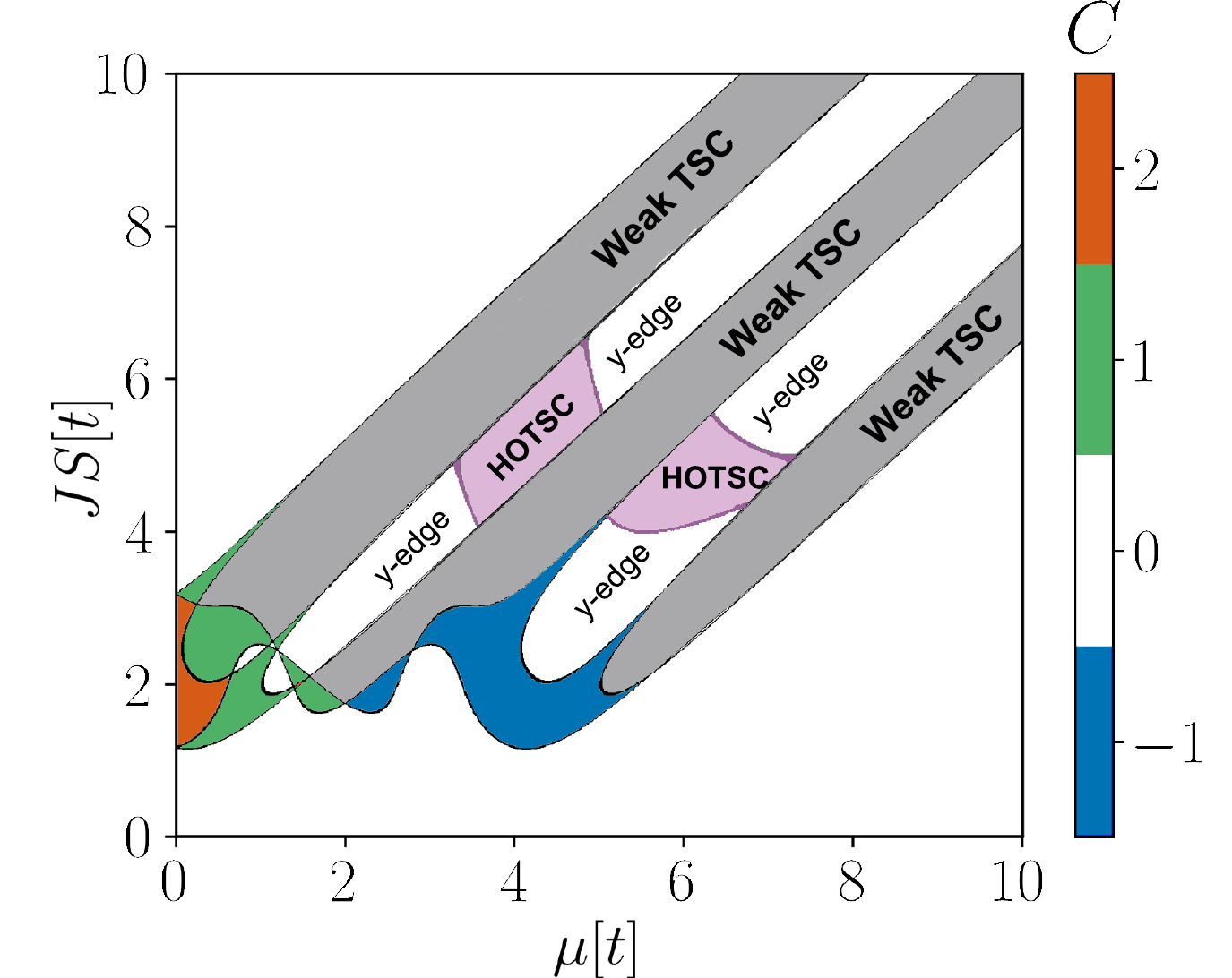}}
    \caption[]{{\bf Topological phase diagrams for different magnetic stacking configurations} Topological phase diagrams for MSH systems as a function of $\mu$ and $JS$ with {\bf a} a $(J, 0)$ stacking, {\bf b} a $(J, 0, 0, J)$ stacking, and {\bf c} a $(J,J,J, 0, 0)$ stacking. Parameters are $\lambda=0.8t$ and $\Delta=1.2t$ for {\bf a}, {\bf b} and $\lambda=0.8t$ and $\Delta=0.8t$ for {\bf c}. The system size $(N_i)$ and the island size $(W_i)$ used for calculating the $y$-edge and $x$-edge gap closings are $(N_x,W_x)=(120,80)$ and $(N_y,W_y)=(280,240)$ respectively.}
    \label{fig:3}
\end{figure*}

\noindent {\bf Local density of states and edge spectral functions\\}
To further characterize the nature of the topological phases identified in Figs.~\ref{fig:1c} and~\ref{fig:1d}, we consider a finite island of magnetic adatom chains with $(J,0,J)$ stacking on the surface of the $s$-wave superconductor. In Figs.~\figref{2}{\bf{a}}-{\bf{c}} we show the zero-energy local density of states (LDOS) for characteristic parameter values in the (strong) $C=-1$, weak, and higher order topological phase. As expected from the bulk-boundary correspondence, we find that the strong $C=1$ phase exhibits zero-energy modes on all edges of the island. In contrast, in the weak topological phase, zero-energy states exists only along the $y$-edges, in agreement with $v_{y,\pi}$ being nonzero and $C=0$. Finally, in the HOTSC phase, the LDOS exhibits four zero-energy states localized on the corners of the island. Since corner modes are characteristic features of a HOTSC, the LDOS presented in Fig.~\figref{2}{\bf c} confirms that the phase bounded by the edge gap closings in Fig.~\ref{fig:1c} is indeed a higher order topological phase.

The characteristic features of these three topological phases are also reflected in the edge spectral functions of MSH systems with only $y$-edges (Figs.~\figref{2}{\bf d}-{\bf f}) and only $x$-edges (Figs.~\figref{2}{\bf g}-{\bf i}). The strong topological phase exhibits $|C|$ chiral Majorana edge modes on both edge types, as shown in Figs.~\figref{2}{\bf d} and {\bf g}. In contrast, in the weak topological phase, a low-energy mid-gap edge state exists only on $y$-edges, as expected from the non-trivial $v_{y,\pi}$ invariant (see Figs.~\figref{2}{\bf e} and {\bf h}). Furthermore, in Fig.\figref{2}{\bf e} we observe that the weak topological edge states are decoupled from the bulk bands and do not exhibit spectral flow between the occupied and unoccupied bulk bands, in contrast to the edge states in the strong topological phases. Finally, in the HOTSC phase, corner modes appear only in systems that are bounded in both in the $x$- and $y$-directions, as shown in Fig.~\figref{2}{\bf c}. As such, systems that are periodic in the $y$- or $x$-directions exhibit a gap in the edge spectral function and no zero-energy modes, as shown in Figs.~\figref{2}{\bf f} and {\bf i}.\\

\noindent {\bf Origin of weak and higher order topological phases\\}
We can understand the emergence of weak topological and higher order topological phases by first identifying the phases of a single pair of adjacent chains of magnetic adatoms, i.e., the topological building blocks of our 2D MSH system. In the relevant regions of the phase diagram, either each chain in the pair individually realizes the topological phase of a Kitaev chain~\cite{Kitaev2001}, or else the two chains can combine together to form a single, effective topological Kitaev chain (see Supplementary Notes 2 and 3). In the latter case, the effective Kitaev chains are uniformly coupled throughout the MSH structure in Fig.~\ref{fig:1a} and form a weak topological phase. The MZM end modes on the effective Kitaev chains are hybridized by the uniform coupling and generate a dispersive Majorana edge band along the $y$-direction, as shown in Figs.~\figref{2}{\bf b} and {\bf e}.

In contrast, the HOTSC phase is realized when each of the adatom chains individually form a topological Kitaev chain. To be explicit, let us first consider the $(J,0,J)$ stacking.  Stacking these chains in this configuration leads to two different couplings of the Kitaev chains: an intra-pair coupling between the two adatom chains in each adjacent pair, and a weaker inter-pair coupling between chains separated by a bare row of substrate. As a result, the MZMs at the ends of each pair of chains are gapped out by the intra-pair coupling, leaving gapped $y$-edges. For the $(J,0,J)$ stacking, this leaves unpaired Kitaev chains on the top and bottom $x$-edges. The MZM end modes from these unpaired chains form the corner modes of the HOTSC phase shown in Fig.~\figref{2}{\bf c}. Alternatively, these corner modes can be understood as arising from topological domain walls between the $x$- and $y$-edges (see Supplementary Note 4).

Using this picture we can also understand the edge transitions for the $(J,0,J)$ stacking phase diagram. We recall that the HOTSC phase is separated from a trivial phase by a gap closing either along the $x$- or $y$-edges (see Fig.~\ref{fig:1c}). The $x$-edge gap closing corresponds to a phase transition of the individual chains from the topological Kitaev chain phase to the trivial phase (see Supplementary Note 2). Consequently, the Majorana corner modes at the ends of the top and bottom chains delocalize horizontally across the x-edge and annihilate at such a phase transition. In contrast, the $y$-edge transition occurs when the intra- and inter- couplings become comparable, allowing the corner modes to delocalize and annihilate along the $y$-edge. We note that the $x$- and $y$-edge transition lines bounding the HOTSC phase terminate at phase transition lines bounding strong topological phases, in which the gap closes simultaneously along the $x$- and $y$-edges.

To understand the effects of the boundary termination on the HOTSC phase (c.f. Figs.~\ref{fig:1c} and ~\ref{fig:1d}), we first note that only for MSH systems with an $x$-edge do the $(J, 0, J)$ and $(J, J, 0)$ stackings lead to systems with a different magnetic structure.
As a result, transition lines in the phase diagram arising from $y$-edge gap closings are the same for $(J, 0, J)$ and $(J, J, 0)$ stackings.
Moreover, when converting an MSH system with a $(J,0,J)$ stacking into one with a $(J, J, 0)$ stacking by adding a single chain of magnetic adatoms to the top and bottom edges of the system, the corner MZMs of the HOTSC phase hybridize with the MZMs of the added chains and annihilate. Hence, the regions of the phase diagram where the system with $(J, 0, J)$ stacking is in the HOTSC phase are trivial for a system with $(J, J, 0)$ stacking, in agreement with the results shown in Figs.~\ref{fig:1c} and ~\ref{fig:1d}.
Conversely, in the trivial phase above the $y$-edge transition of a system with $(J,0,J)$ stacking (see Fig.~\ref{fig:1}{\bf c}), the addition of a single chain at the top and bottom edges, which transforms it into a $(J, J, 0)$ stacked system, leads to the emergence of four corner modes, and a HOTSC phase. The pattern of trivial and topological phases for the two types of terminations are thus complementary across the $y$-edge gap closing line.

We finally note that the sensitivity of the HOTSC phase to the specific termination of the MSH system opens an unprecedented opportunity tune the system between a HOTSC phase and a trivial phase by adding or removing chains of magnetic adatoms using atomic manipulation techniques. \\

\noindent {\bf Systematics of chain stacking and the phase diagram}\\
The results presented above allow us to systematically address the relation between the form of the chain stacking and the resulting topological phase diagram. As previously shown, a superconducting surface fully covered by adatoms results in strong topological superconducting phases, but no weak or higher order topological phase~\cite{Ron2015,Li2016,Rachel2017}. The next-simplest stacking is one that alternates between adatom chains and bare rows of substrate, which we refer to as $(J, 0)$ stacking. The phase diagram for this stacking (see Fig.~\ref{fig:3a}) reveals that this system possesses strong and weak topological phases, but no higher order topological phase.  In the region of the phase diagram where the weak topological phase is located, isolated adatom chains realize
topological Kitaev chains. Since the inter-chain coupling in the $(J,0)$ stacking configuration is uniform, the coupled Kitaev chains generate the weak topological phase, and the boundary MZMs hybridize into a dispersive, mid-gap band.

We generically do not expect to find higher order topology in stackings with a single adatom per unit cell. Such stackings do not allow for spatially-modulated weak-strong couplings between adatom chains, and therefore cannot give rise to a HOTSC phase. Specifically, the adatom chains in such configurations are always separated from each other by a fixed number of substrate rows, and cannot form the dimerized pattern that is crucial to the emergence of the HOTSC phase discussed above. As a result, we find that MSH systems with any stacking with a single adatom per unit cell, such as $(J, 0)$, $(J, 0, 0)$, $(J, 0, 0, 0),$ etc., do not possess higher order topological phases.

In contrast, MSH systems with stackings such as $(J, 0, 0, J)$ possess a phase diagram that is qualitatively similar to that of the $(J, 0, J)$ stacking, but with additional edge gap closings and several regions containing an extrinsic HOTSC phase (see Fig.~\ref{fig:3b}). When increasing the size of the unit cell of the stacking even further, for example, by considering a $(J,J,J,0, 0)$ stacking, the phase diagram (see Fig.~\ref{fig:3c}), while still exhibiting the same strong, weak, and high-order topological phases, becomes more complex and the number of disjoint topological regions increases.\\

\begin{figure*}
    \centering
    \hypertarget{fig:4}{}
    \includegraphics[width=0.8\textwidth]{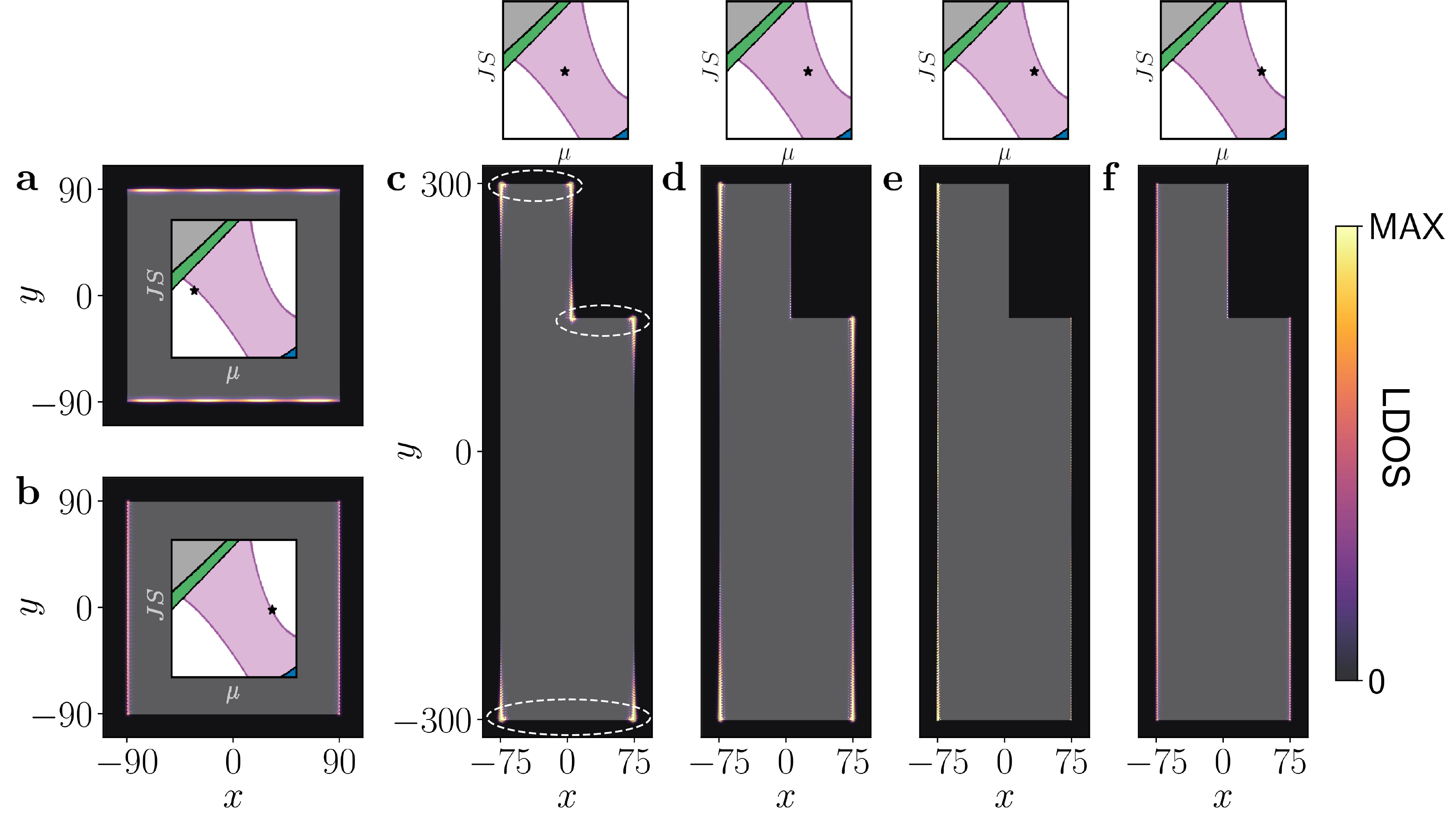}
    \caption[]{{\bf Majorana qubit generation}
    The zero-energy LDOS of a square MHS island demonstrating the delocalization of corner Majorana modes along {\bf a} the $x$-edges  at the $x$-edge gap closing line ($\mu=2.4t$), and {\bf b} the $y$-edges at the $y$-edge gap closing line ($\mu=3.2t$), as indicated by the stars in the inset phase diagrams. {\bf c}-{\bf f} Zero-energy LDOS of an armchair MSH system with six corners, plotted for varying chemical potential $\mu$ (indicated by stars in the phase diagrams in the upper row). Black regions indicate the bare superconducting substrate and the white dashed ovals emphasize the three pairs of Majorana corner modes. Parameters for all plots are $(\lambda,\Delta,JS)=(0.8, 1.2, 3)t$.}
    \label{fig:4}
\end{figure*}

\noindent {\bf Majorana Qubit Generation and Fusion}\\
One of the key technological applications of topological superconductors is topological quantum computing, which exploits the robust, non-local information storage capacity of MZMs~\cite{Nayak2008}. The topological quantum computation paradigm is built upon the non-trivial MZM fusion rules (two MZMs fuse to either an un-occupied/vacuum channel or an occupied/quasiparticle channel) that must be verified in any topological qubit platform~\cite{aasen2016milestones,alicea2011non}. Additionally, in the original proposals for topological quantum computing, quantum operations on MZM qubits were proposed to be carried out by physically braiding the MZMs. However, it was subsequently shown that measurement-based quantum computation \cite{Bonderson2008, Bonderson2009} can generate non-trivial unitary operations on the MZM qubit space by tuning the relative coupling strengths between MZMs and performing charge read-out measurements.

The extrinsic HOTSC phase we predict here provides an advantageous platform for generating, coupling, and fusing MZMs. In our platform, MZM qubits are geometrically defined by the magnetic adatom configuration. Engineering the shape of the region containing magnetic adatoms determines the locations and relative coupling strengths of the MZMs. For instance, in Fig. \ref{fig:4} we show examples of an MSH island with a square geometry that harbors four MZMs (a tetron) and an arm-chair geometry that hosts six MZMs (a hexon)~\cite{karzig_scalable_2017}. By carefully choosing the configurations and edge-lengths between MZMs, we can engineer pair-wise tuning of MZM coupling and fusion simply by adjusting the global electrostatic potential. For example, consider a $(J,0,J)$ stacked square geometry with $\mu$ chosen deep in the HOTSC phase. Tuning $\mu$ toward an edge transition point then causes the MZMs to delocalize and pair-wise couple. In fact, choosing which transition point to approach determines whether the MZMs become more strongly coupled to their horizontal or vertical neighbors. As one of these transition points is approached, the coupling between MZMs on the ends of the associated edge strengthens, and eventually the corner MZMs fuse along the edge. It is therefore possible to selectively couple and fuse MZMs with electrostatic gate control.

Here we provide an explicit example of how one can verify the MZM fusion rules using the square/tetron geometry (c.f. Figs.~\figref[a]{4}-\figref[b]{4})~\cite{aasen2016milestones}. We start in the HOTSC phase, but with $\mu$ very near an edge transition. If we tune $\mu$ away from the transition point, then two pairs of MZMs, $(\gamma_1, \gamma_2)$ and $(\gamma_3, \gamma_4)$, are nucleated from the vacuum. In a control experiment, these MZMs can be re-fused trivially to the vacuum channel, i.e., $\gamma_1$ fuses with $\gamma_2,$ and $\gamma_3$ fuses with $\gamma_4$, by returning to the initial transition point. In the non-trivial experiment, after the initial nucleation from the vacuum, MZMs from different pairs are fused together, e.g., $\gamma_1$ with $\gamma_3$ and $\gamma_2$ with $\gamma_4,$ by approaching the other edge transition. In this case, the vacuum and quasi-particle fusion outcomes are produced with equal probability. The result of the fusion process can then be determined by measuring the charge of the final state, for example through a capacitively coupled quantum dot~\cite{aasen2016milestones}.

The atomically precise control afforded by atomic manipulation techniques in the fabrication of MSH systems makes it possible to pattern arbitrary geometries of coupled tetron and hexon devices. Furthermore, using geometries with edges of varying lengths, we can implement more complex MZM coupling operations. For example, in the hexon geometry of Fig.~\ref{fig:4}{\bf c}, the lengths of the horizontal and vertical edges are chosen to permit the fusion of some pairs of MZMs independently of other pairs. Specifically, in Fig.~\ref{fig:4}{\bf c} we consider an MSH system tuned into the middle of the HOTSC phase and possessing three pairs of corner MZMs enclosed by the dashed white ovals. As demonstrated in Figs.~~\ref{fig:4}{\bf c}-{\bf f}, tuning $\mu$ towards the $y$-edge gap closing performs three sequential fusions along the three $y$-edges in order of increasing edge length. Remarkably, the freedom afforded by geometric patterning allows tetron or hexon fusion operations to be implemented by tuning a single electrostatic gate.\\[-0.3cm]

\noindent {\bf Discussion}\\
We have demonstrated that MSH systems with a stacked magnetic structure give rise to HOTSC phases. The HOTSC we identified is a boundary obstructed phase that is separated from the trivial phase by edge, rather than bulk, gap closings. In addition, stacked MSH systems also possess strong and weak topological superconducting phases. We have shown that all three types of topological phases possess characteristic spectroscopic signatures for finite size magnetic islands, which can be employed to distinguish them using scanning tunneling spectroscopy. In particular, an island in the HOTSC phase exhibits corner modes, i.e., MZMs that are distinctly localized on the corners of the island. Moreover, we have shown that the HOTSC phase is sensitive to the particular magnetic stacking of the MSH system: in certain regions of the phase diagram where the system is a HOTSC for $(J,0,J)$ stacking, it is trivial for a $(J,J,0)$ stacking. This opens the unprecedented ability to tune MSH systems between trivial and topological phases using atomic manipulation techniques, as a MSH system with $(J,0,J)$ stacking can be transformed into one with $(J,J,0)$ stacking by adding a single magnetic chain to its top and bottom edges. By identifying the microscopic mechanism responsible for the emergence of HOTSC and weak topological phases in MSH systems, i.e., the interplay of intra- and inter-pair couplings in the stacked system, we demonstrated that these phases are generic features of stacked magnetic structures. The unique structure of Majorana corner modes in the HOTSC phase also provide a new path for the realization of topological quantum gates through Majorana fusion.

While previous proposals for the creation of higher order topological phases have required either complex band structures or complex pairing \cite{wang2019,Zhang2019,yan2019higher,ahn2020higher}, our results show that such phases can be readily engineered using the standard features of MSH systems: magnetic adatoms that can be arranged in atomically precise structures using atomic manipulation techniques, a Rashba spin-orbit interaction arising from the broken inversion symmetry on surfaces, and a hard $s$-wave superconducting gap. As such, MSH systems provide a versatile platform for the study of elusive higher order topological phases and their use in the implementation of topological quantum gates. Our findings also raise the interesting question of whether higher order topological phases can be created on lattices with different spatial symmetries, such as triangular lattices, which are also realized in MSH systems~\cite{Palacio-Morales2019,Bazarnik2022}. \\

\noindent {\bf Methods} \\
To compute the location of the bulk gap closings, we diagonalize the Hamiltonian in momentum space. If for any momentum in the Brillouin zone (BZ) the lowest positive energy state possesses an energy $E<0.001t$, we consider the system to be gapless and take the corresponding values of $\mu$ and $JS$ to represent a bulk gap closing point in the phase diagram. Similarly, to compute the location of $x$- or $y$-edge gap closings we diagonalize the Hamiltonian in a cylindrical geometry with periodic boundary conditions along one direction and open boundary conditions along the other. The open boundary direction spans $N$ sites with a magnetic island of width $W$ deposited in the middle. If for any momentum in the 1D BZ the lowest positive energy state possesses an energy $E<0.001t$, we consider the system to be gapless, and take the corresponding values of $\mu$ and  $JS$ to represent an  edge gap closing point in the phase diagram (unless this coincide with a bulk gap closing). The Chern numbers in the phase diagram are computed using Eq.~\eqref{eq:C}. The zero energy LDOS is computed by first casting the Hamiltonian of Eq.~\eqref{eq:H} into a matrix form, $\hat{H}$, in real space and Nambu space. Then, we compute the retarded Green's function matrix, $g^{r}(\omega)=[(\omega +i\delta)\hat{1} -\hat{H}]^{-1}$, from which the zero energy LDOS can be obtained via $\rho(\vb{r})=-\Im g^{r}_{\vb{r}\vb{r}}(\omega=0)/\pi$. Similarly, the $i$-edge spectral function can be obtained via $A(k_i,E)=-\Im g^{r}_{xx}(k_i,E)/\pi$, where $x$ is the coordinate of the $i$-edge and $g^{r}(k_i,E)$ is the retarded Green's function matrix computed using the Hamiltonian in the aforementioned cylindrical geometry periodic along the $i$-direction. Both LDOS and spectral functions are summed over spin and Nambu degrees of freedom.\\

\noindent {\bf Acknowledgements}\\
This study was supported by the Center for Quantum Sensing and Quantum Materials, an Energy Frontier Research Center funded by the U. S. Department of Energy, Office of Science, Basic Energy Sciences under Award DE-SC0021238.\\

\noindent {\bf Author contributions} \\
B.B., T.L.H., and D.K.M. conceived and supervised the project. K.H.W., M.R.H, J.G, and A.M. performed the theoretical calculations. All authors discussed the results. M.R.H., J.G., T.L.H., and D.K.M. wrote the
manuscript with contributions from all the authors.\\

\noindent {\bf Competing interests}\\
The authors declare no competing interests.\\

\noindent {\bf Data availability}: Original data available from the authors on reasonable
request.\\

\noindent {\bf Code availability}: The codes that were employed in this study are available from the authors upon reasonable request.\\


\begin{thebibliography}{10}
\expandafter\ifx\csname url\endcsname\relax
  \def\url#1{\texttt{#1}}\fi
\expandafter\ifx\csname urlprefix\endcsname\relax\def\urlprefix{URL }\fi
\providecommand{\bibinfo}[2]{#2}
\providecommand{\eprint}[2][]{\url{#2}}

\bibitem{Nayak2008}
\bibinfo{author}{Nayak, C.}, \bibinfo{author}{Simon, S.~H.},
  \bibinfo{author}{Stern, A.}, \bibinfo{author}{Freedman, M.} \&
  \bibinfo{author}{Das~Sarma, S.}
\newblock \bibinfo{title}{Non-abelian anyons and topological quantum
  computation}.
\newblock \emph{\bibinfo{journal}{Rev. Mod. Phys.}}
  \textbf{\bibinfo{volume}{80}}, \bibinfo{pages}{1083--1159}
  (\bibinfo{year}{2008}).
\newblock \urlprefix\url{https://link.aps.org/doi/10.1103/RevModPhys.80.1083}.

\bibitem{Nadj-Perge2014}
\bibinfo{author}{{Nadj-Perge}, S.} \emph{et~al.}
\newblock \bibinfo{title}{Observation of {{Majorana}} fermions in ferromagnetic
  atomic chains on a superconductor}.
\newblock \emph{\bibinfo{journal}{Science}} \textbf{\bibinfo{volume}{346}},
  \bibinfo{pages}{602--607} (\bibinfo{year}{2014}).

\bibitem{Ruby2015}
\bibinfo{author}{Ruby, M.} \emph{et~al.}
\newblock \bibinfo{title}{End states and subgap structure in proximity-coupled
  chains of magnetic adatoms}.
\newblock \emph{\bibinfo{journal}{Phys. Rev. Lett.}}
  \textbf{\bibinfo{volume}{115}}, \bibinfo{pages}{197204}
  (\bibinfo{year}{2015}).

\bibitem{Pawlak2016}
\bibinfo{author}{Pawlak, R.} \emph{et~al.}
\newblock \bibinfo{title}{Probing atomic structure and {{Majorana}}
  wavefunctions in mono-atomic {{Fe}} chains on superconducting {{Pb}}
  surface}.
\newblock \emph{\bibinfo{journal}{npj Quantum Inf.}}
  \textbf{\bibinfo{volume}{2}}, \bibinfo{pages}{16035} (\bibinfo{year}{2016}).

\bibitem{Kim2018}
\bibinfo{author}{Kim, H.} \emph{et~al.}
\newblock \bibinfo{title}{Toward tailoring {{Majorana}} bound states in
  artificially constructed magnetic atom chains on elemental superconductors}.
\newblock \emph{\bibinfo{journal}{Sci. Adv.}} \textbf{\bibinfo{volume}{4}},
  \bibinfo{pages}{eaar5251} (\bibinfo{year}{2018}).
\newblock \urlprefix\url{https://doi.org/10.1126/sciadv.aar5251}.

\bibitem{Menard2017}
\bibinfo{author}{M{\'e}nard, G.~C.} \emph{et~al.}
\newblock \bibinfo{title}{Two-dimensional topological superconductivity in
  {{Pb}}/{{Co}}/{{Si}}(111)}.
\newblock \emph{\bibinfo{journal}{Nat. Commun.}} \textbf{\bibinfo{volume}{8}},
  \bibinfo{pages}{2040} (\bibinfo{year}{2017}).
\newblock \urlprefix\url{https://doi.org/10.1038/s41467-017-02192-x}.

\bibitem{Palacio-Morales2019}
\bibinfo{author}{{Palacio-Morales}, A.} \emph{et~al.}
\newblock \bibinfo{title}{Atomic-scale interface engineering of {{Majorana}}
  edge modes in a {{2D}} magnet-superconductor hybrid system}.
\newblock \emph{\bibinfo{journal}{Sci. Adv.}} \textbf{\bibinfo{volume}{5}},
  \bibinfo{pages}{eaav6600} (\bibinfo{year}{2019}).
\newblock \urlprefix\url{https://doi.org/10.1126/sciadv.aav6600}.

\bibitem{Kezilebieke2020}
\bibinfo{author}{Kezilebieke, S.} \emph{et~al.}
\newblock \bibinfo{title}{Topological superconductivity in a van der {{Waals}}
  heterostructure}.
\newblock \emph{\bibinfo{journal}{Nature}} \textbf{\bibinfo{volume}{588}},
  \bibinfo{pages}{424--428} (\bibinfo{year}{2020}).

\bibitem{Mascot2021}
\bibinfo{author}{Mascot, E.}, \bibinfo{author}{Bedow, J.},
  \bibinfo{author}{Graham, M.}, \bibinfo{author}{Rachel, S.} \&
  \bibinfo{author}{Morr, D.~K.}
\newblock \bibinfo{title}{Topological superconductivity in skyrmion lattices}.
\newblock \emph{\bibinfo{journal}{npj Quantum Mater.}}
  \textbf{\bibinfo{volume}{6}}, \bibinfo{pages}{6} (\bibinfo{year}{2021}).
\newblock \urlprefix\url{https://doi.org/10.1038/s41535-020-00299-x}.

\bibitem{Bedow2020}
\bibinfo{author}{Bedow, J.} \emph{et~al.}
\newblock \bibinfo{title}{Topological superconductivity induced by a triple-
  {\textbf{q}} magnetic structure}.
\newblock \emph{\bibinfo{journal}{Phys. Rev. B}}
  \textbf{\bibinfo{volume}{102}}, \bibinfo{pages}{180504}
  (\bibinfo{year}{2020}).
\newblock \urlprefix\url{https://doi.org/10.1103/PhysRevB.102.180504}.

\bibitem{Bazarnik2022}
\bibinfo{author}{Bazarnik, M.} \emph{et~al.}
\newblock \bibinfo{title}{Antiferromagnetism-driven two-dimensional topological
  nodal-point superconductivity}.
\newblock \emph{\bibinfo{journal}{arXiv:2208.12018}}  (\bibinfo{year}{2022}).
\newblock \urlprefix\url{https://arxiv.org/abs/2208.12018}.

\bibitem{wang2019}
\bibinfo{author}{Wang, Y.}, \bibinfo{author}{Lin, M.} \&
  \bibinfo{author}{Hughes, T.~L.}
\newblock \bibinfo{title}{Weak-pairing higher order topological
  superconductors}.
\newblock \emph{\bibinfo{journal}{Phys. Rev. B}} \textbf{\bibinfo{volume}{98}},
  \bibinfo{pages}{165144} (\bibinfo{year}{2018}).
\newblock \urlprefix\url{https://link.aps.org/doi/10.1103/PhysRevB.98.165144}.

\bibitem{khalaf2018higher}
\bibinfo{author}{Khalaf, E.}
\newblock \bibinfo{title}{Higher-order topological insulators and
  superconductors protected by inversion symmetry}.
\newblock \emph{\bibinfo{journal}{Physical Review B}}
  \textbf{\bibinfo{volume}{97}}, \bibinfo{pages}{205136}
  (\bibinfo{year}{2018}).

\bibitem{ono2020refined}
\bibinfo{author}{Ono, S.}, \bibinfo{author}{Po, H.~C.} \&
  \bibinfo{author}{Watanabe, H.}
\newblock \bibinfo{title}{Refined symmetry indicators for topological
  superconductors in all space groups}.
\newblock \emph{\bibinfo{journal}{Science advances}}
  \textbf{\bibinfo{volume}{6}}, \bibinfo{pages}{eaaz8367}
  (\bibinfo{year}{2020}).

\bibitem{ono2021z}
\bibinfo{author}{Ono, S.}, \bibinfo{author}{Po, H.~C.} \&
  \bibinfo{author}{Shiozaki, K.}
\newblock \bibinfo{title}{Z 2-enriched symmetry indicators for topological
  superconductors in the 1651 magnetic space groups}.
\newblock \emph{\bibinfo{journal}{Physical Review Research}}
  \textbf{\bibinfo{volume}{3}}, \bibinfo{pages}{023086} (\bibinfo{year}{2021}).

\bibitem{schindler2020pairing}
\bibinfo{author}{Schindler, F.}, \bibinfo{author}{Bradlyn, B.},
  \bibinfo{author}{Fischer, M.~H.} \& \bibinfo{author}{Neupert, T.}
\newblock \bibinfo{title}{Pairing obstructions in topological superconductors}.
\newblock \emph{\bibinfo{journal}{Physical review letters}}
  \textbf{\bibinfo{volume}{124}}, \bibinfo{pages}{247001}
  (\bibinfo{year}{2020}).

\bibitem{tang2022high}
\bibinfo{author}{Tang, F.}, \bibinfo{author}{Ono, S.}, \bibinfo{author}{Wan,
  X.} \& \bibinfo{author}{Watanabe, H.}
\newblock \bibinfo{title}{High-throughput investigations of topological and
  nodal superconductors}.
\newblock \emph{\bibinfo{journal}{Physical Review Letters}}
  \textbf{\bibinfo{volume}{129}}, \bibinfo{pages}{027001}
  (\bibinfo{year}{2022}).

\bibitem{langbehn2017reflection}
\bibinfo{author}{Langbehn, J.}, \bibinfo{author}{Peng, Y.},
  \bibinfo{author}{Trifunovic, L.}, \bibinfo{author}{von Oppen, F.} \&
  \bibinfo{author}{Brouwer, P.~W.}
\newblock \bibinfo{title}{Reflection-symmetric second-order topological
  insulators and superconductors}.
\newblock \emph{\bibinfo{journal}{Physical review letters}}
  \textbf{\bibinfo{volume}{119}}, \bibinfo{pages}{246401}
  (\bibinfo{year}{2017}).

\bibitem{hsu2018majorana}
\bibinfo{author}{Hsu, C.-H.}, \bibinfo{author}{Stano, P.},
  \bibinfo{author}{Klinovaja, J.} \& \bibinfo{author}{Loss, D.}
\newblock \bibinfo{title}{Majorana kramers pairs in higher-order topological
  insulators}.
\newblock \emph{\bibinfo{journal}{Physical review letters}}
  \textbf{\bibinfo{volume}{121}}, \bibinfo{pages}{196801}
  (\bibinfo{year}{2018}).

\bibitem{benalcazar2017science}
\bibinfo{author}{Benalcazar, W.~A.}, \bibinfo{author}{Bernevig, B.~A.} \&
  \bibinfo{author}{Hughes, T.~L.}
\newblock \bibinfo{title}{{Quantized electric multipole insulators}}.
\newblock \emph{\bibinfo{journal}{Science}} \textbf{\bibinfo{volume}{357}},
  \bibinfo{pages}{61--66} (\bibinfo{year}{2017}).
\newblock \urlprefix\url{https://doi.org/10.1126/science.aah6442}.
\newblock \eprint{1611.07987}.

\bibitem{benalcazar2017prb}
\bibinfo{author}{Benalcazar, W.~A.}, \bibinfo{author}{Bernevig, B.~A.} \&
  \bibinfo{author}{Hughes, T.~L.}
\newblock \bibinfo{title}{Electric multipole moments, topological multipole
  moment pumping, and chiral hinge states in crystalline insulators}.
\newblock \emph{\bibinfo{journal}{Phys. Rev. B}} \textbf{\bibinfo{volume}{96}},
  \bibinfo{pages}{245115} (\bibinfo{year}{2017}).
\newblock \urlprefix\url{https://link.aps.org/doi/10.1103/PhysRevB.96.245115}.

\bibitem{Geier2018}
\bibinfo{author}{Geier, M.}, \bibinfo{author}{Trifunovic, L.},
  \bibinfo{author}{Hoskam, M.} \& \bibinfo{author}{Brouwer, P.~W.}
\newblock \bibinfo{title}{Second-order topological insulators and
  superconductors with an order-two crystalline symmetry}.
\newblock \emph{\bibinfo{journal}{Phys. Rev. B}} \textbf{\bibinfo{volume}{97}},
  \bibinfo{pages}{205135} (\bibinfo{year}{2018}).

\bibitem{Khalaf2021}
\bibinfo{author}{Khalaf, E.}, \bibinfo{author}{Benalcazar, W.~A.},
  \bibinfo{author}{Hughes, T.~L.} \& \bibinfo{author}{Queiroz, R.}
\newblock \bibinfo{title}{Boundary-obstructed topological phases}.
\newblock \emph{\bibinfo{journal}{Phys. Rev. Research}}
  \textbf{\bibinfo{volume}{3}}, \bibinfo{pages}{013239} (\bibinfo{year}{2021}).
\newblock
  \urlprefix\url{https://link.aps.org/doi/10.1103/PhysRevResearch.3.013239}.

\bibitem{Ron2015}
\bibinfo{author}{R{\"o}ntynen, J.} \& \bibinfo{author}{Ojanen, T.}
\newblock \bibinfo{title}{Topological superconductivity and high chern numbers
  in {{2D}} ferromagnetic {{Shiba}} lattices}.
\newblock \emph{\bibinfo{journal}{Phys. Rev. Lett.}}
  \textbf{\bibinfo{volume}{114}}, \bibinfo{pages}{236803}
  (\bibinfo{year}{2015}).

\bibitem{Li2016}
\bibinfo{author}{Li, J.} \emph{et~al.}
\newblock \bibinfo{title}{Two-dimensional chiral topological superconductivity
  in {{Shiba}} lattices}.
\newblock \emph{\bibinfo{journal}{Nat. Commun.}} \textbf{\bibinfo{volume}{7}},
  \bibinfo{pages}{12297} (\bibinfo{year}{2016}).

\bibitem{Rachel2017}
\bibinfo{author}{Rachel, S.}, \bibinfo{author}{Mascot, E.},
  \bibinfo{author}{Cocklin, S.}, \bibinfo{author}{Vojta, M.} \&
  \bibinfo{author}{Morr, D.~K.}
\newblock \bibinfo{title}{Quantized charge transport in chiral {{Majorana}}
  edge modes}.
\newblock \emph{\bibinfo{journal}{Phys. Rev. B}} \textbf{\bibinfo{volume}{96}},
  \bibinfo{pages}{205131} (\bibinfo{year}{2017}).
\newblock \urlprefix\url{https://doi.org/10.1103/PhysRevB.96.205131}.

\bibitem{Balatsky2006}
\bibinfo{author}{Balatsky, A.~V.}, \bibinfo{author}{Vekhter, I.} \&
  \bibinfo{author}{Zhu, J.-X.}
\newblock \bibinfo{title}{Impurity-induced states in conventional and
  unconventional superconductors}.
\newblock \emph{\bibinfo{journal}{Rev. Mod. Phys.}}
  \textbf{\bibinfo{volume}{78}}, \bibinfo{pages}{373--433}
  (\bibinfo{year}{2006}).

\bibitem{Heinrich2018}
\bibinfo{author}{Heinrich, B.~W.}, \bibinfo{author}{Pascual, J.~I.} \&
  \bibinfo{author}{Franke, K.~J.}
\newblock \bibinfo{title}{Single magnetic adsorbates on s -wave
  superconductors}.
\newblock \emph{\bibinfo{journal}{Progress in Surface Science}}
  \textbf{\bibinfo{volume}{93}}, \bibinfo{pages}{1--19} (\bibinfo{year}{2018}).

\bibitem{altlandzirnbauer}
\bibinfo{author}{Altland, A.} \& \bibinfo{author}{Zirnbauer, M.~R.}
\newblock \bibinfo{title}{Nonstandard symmetry classes in mesoscopic
  normal-superconducting hybrid structures}.
\newblock \emph{\bibinfo{journal}{Phys. Rev. B}} \textbf{\bibinfo{volume}{55}},
  \bibinfo{pages}{1142--1161} (\bibinfo{year}{1997}).
\newblock \urlprefix\url{https://link.aps.org/doi/10.1103/PhysRevB.55.1142}.

\bibitem{kitaev2009periodic}
\bibinfo{author}{Kitaev, A.}
\newblock \bibinfo{title}{Periodic table for topological insulators and
  superconductors}.
\newblock In \emph{\bibinfo{booktitle}{AIP conference proceedings}}, vol.
  \bibinfo{volume}{1134}, \bibinfo{pages}{22--30}
  (\bibinfo{organization}{American Institute of Physics},
  \bibinfo{year}{2009}).
\newblock \urlprefix\url{https://doi.org/10.1063/1.3149495}.

\bibitem{ryu2010topological}
\bibinfo{author}{Ryu, S.}, \bibinfo{author}{Schnyder, A.~P.},
  \bibinfo{author}{Furusaki, A.} \& \bibinfo{author}{Ludwig, A.~W.}
\newblock \bibinfo{title}{Topological insulators and superconductors: tenfold
  way and dimensional hierarchy}.
\newblock \emph{\bibinfo{journal}{New Journal of Physics}}
  \textbf{\bibinfo{volume}{12}}, \bibinfo{pages}{065010}
  (\bibinfo{year}{2010}).
\newblock \urlprefix\url{https://doi.org/10.1088/1367-2630/12/6/065010}.

\bibitem{avron1983homotopy}
\bibinfo{author}{Avron, J.~E.}, \bibinfo{author}{Seiler, R.} \&
  \bibinfo{author}{Simon, B.}
\newblock \bibinfo{title}{Homotopy and quantization in condensed matter
  physics}.
\newblock \emph{\bibinfo{journal}{Physical review letters}}
  \textbf{\bibinfo{volume}{51}}, \bibinfo{pages}{51} (\bibinfo{year}{1983}).
\newblock \urlprefix\url{https://doi.org/10.1103/PhysRevLett.51.51}.

\bibitem{seroussi2014topological}
\bibinfo{author}{Seroussi, I.}, \bibinfo{author}{Berg, E.} \&
  \bibinfo{author}{Oreg, Y.}
\newblock \bibinfo{title}{Topological superconducting phases of weakly coupled
  quantum wires}.
\newblock \emph{\bibinfo{journal}{Physical Review B}}
  \textbf{\bibinfo{volume}{89}}, \bibinfo{pages}{104523}
  (\bibinfo{year}{2014}).
\newblock \urlprefix\url{https://doi.org/10.1103/PhysRevB.89.104523}.

\bibitem{fu2007topological}
\bibinfo{author}{Fu, L.} \& \bibinfo{author}{Kane, C.~L.}
\newblock \bibinfo{title}{Topological insulators with inversion symmetry}.
\newblock \emph{\bibinfo{journal}{Physical Review B}}
  \textbf{\bibinfo{volume}{76}}, \bibinfo{pages}{045302}
  (\bibinfo{year}{2007}).
\newblock \urlprefix\url{https://doi.org/10.1103/PhysRevB.76.045302}.

\bibitem{Kitaev2001}
\bibinfo{author}{Kitaev, A.~Y.}
\newblock \bibinfo{title}{{Unpaired Majorana fermions in quantum wires}}.
\newblock \emph{\bibinfo{journal}{Phys. Usp.}} \textbf{\bibinfo{volume}{44}},
  \bibinfo{pages}{131--136} (\bibinfo{year}{2001}).
\newblock \urlprefix\url{https://doi.org/10.1070/1063-7869/44/10S/S29}.
\newblock \eprint{cond-mat/0010440}.

\bibitem{aasen2016milestones}
\bibinfo{author}{Aasen, D.} \emph{et~al.}
\newblock \bibinfo{title}{Milestones toward majorana-based quantum computing}.
\newblock \emph{\bibinfo{journal}{Physical Review X}}
  \textbf{\bibinfo{volume}{6}}, \bibinfo{pages}{031016} (\bibinfo{year}{2016}).
\newblock \urlprefix\url{https://doi.org/10.1103/PhysRevX.6.031016}.

\bibitem{alicea2011non}
\bibinfo{author}{Alicea, J.}, \bibinfo{author}{Oreg, Y.},
  \bibinfo{author}{Refael, G.}, \bibinfo{author}{Von~Oppen, F.} \&
  \bibinfo{author}{Fisher, M.}
\newblock \bibinfo{title}{Non-abelian statistics and topological quantum
  information processing in 1d wire networks}.
\newblock \emph{\bibinfo{journal}{Nature Physics}}
  \textbf{\bibinfo{volume}{7}}, \bibinfo{pages}{412--417}
  (\bibinfo{year}{2011}).
\newblock \urlprefix\url{https://doi.org/10.1038/nphys1915}.

\bibitem{Bonderson2008}
\bibinfo{author}{Bonderson, P.}, \bibinfo{author}{Freedman, M.} \&
  \bibinfo{author}{Nayak, C.}
\newblock \bibinfo{title}{Measurement-only topological quantum computation}.
\newblock \emph{\bibinfo{journal}{Phys. Rev. Lett.}}
  \textbf{\bibinfo{volume}{101}}, \bibinfo{pages}{010501}
  (\bibinfo{year}{2008}).
\newblock
  \urlprefix\url{https://link.aps.org/doi/10.1103/PhysRevLett.101.010501}.

\bibitem{Bonderson2009}
\bibinfo{author}{Bonderson, P.}, \bibinfo{author}{Freedman, M.} \&
  \bibinfo{author}{Nayak, C.}
\newblock \bibinfo{title}{Measurement-only topological quantum computation via
  anyonic interferometry}.
\newblock \emph{\bibinfo{journal}{Annals of Physics}}
  \textbf{\bibinfo{volume}{324}}, \bibinfo{pages}{787--826}
  (\bibinfo{year}{2009}).
\newblock
  \urlprefix\url{https://www.sciencedirect.com/science/article/pii/S0003491608001577}.

\bibitem{karzig_scalable_2017}
\bibinfo{author}{Karzig, T.} \emph{et~al.}
\newblock \bibinfo{title}{Scalable designs for
  quasiparticle-poisoning-protected topological quantum computation with
  {Majorana} zero modes}.
\newblock \emph{\bibinfo{journal}{Physical Review B}}
  \textbf{\bibinfo{volume}{95}}, \bibinfo{pages}{235305}
  (\bibinfo{year}{2017}).
\newblock \urlprefix\url{http://link.aps.org/doi/10.1103/PhysRevB.95.235305}.

\bibitem{Zhang2019}
\bibinfo{author}{Zhang, R.-X.}, \bibinfo{author}{Cole, W.~S.},
  \bibinfo{author}{Wu, X.} \& \bibinfo{author}{Das~Sarma, S.}
\newblock \bibinfo{title}{Higher-{{Order Topology}} and {{Nodal Topological
  Superconductivity}} in {{Fe}}({{Se}},{{Te}}) {{Heterostructures}}}.
\newblock \emph{\bibinfo{journal}{Phys. Rev. Lett.}}
  \textbf{\bibinfo{volume}{123}}, \bibinfo{pages}{167001}
  (\bibinfo{year}{2019}).

\bibitem{yan2019higher}
\bibinfo{author}{Yan, Z.}
\newblock \bibinfo{title}{Higher-order topological odd-parity superconductors}.
\newblock \emph{\bibinfo{journal}{Physical review letters}}
  \textbf{\bibinfo{volume}{123}}, \bibinfo{pages}{177001}
  (\bibinfo{year}{2019}).

\bibitem{ahn2020higher}
\bibinfo{author}{Ahn, J.} \& \bibinfo{author}{Yang, B.-J.}
\newblock \bibinfo{title}{Higher-order topological superconductivity of
  spin-polarized fermions}.
\newblock \emph{\bibinfo{journal}{Physical Review Research}}
  \textbf{\bibinfo{volume}{2}}, \bibinfo{pages}{012060} (\bibinfo{year}{2020}).

\end{thebibliography}
\end{document}


\renewcommand{\theequation}{S\arabic{equation}}
\renewcommand{\figurename}{Supplementary Figure}

\title{Higher Order Topological Superconductivity in Magnet-Superconductor Hybrid Systems\\ Supplemental Information}

\author{Ka Ho Wong}\affiliation{\UICPHYS}
\author{Mark R. Hirsbrunner}\affiliation{\UIUCPHYS}
\author{Jacopo Gliozzi}\affiliation{\UIUCPHYS}
\author{Arbaz Malik}\affiliation{\UICPHYS}
\author{Barry Bradlyn}\affiliation{\UIUCPHYS}
\author{Taylor L. Hughes}\affiliation{\UIUCPHYS}
\author{Dirk K. Morr}\affiliation{\UICPHYS}

\maketitle

{\bf Supplementary Note 1: Topological Phase diagram for smaller superconducting order parameter}

\begin{figure}[h]
        \centering
        \includegraphics[width=0.45\textwidth]{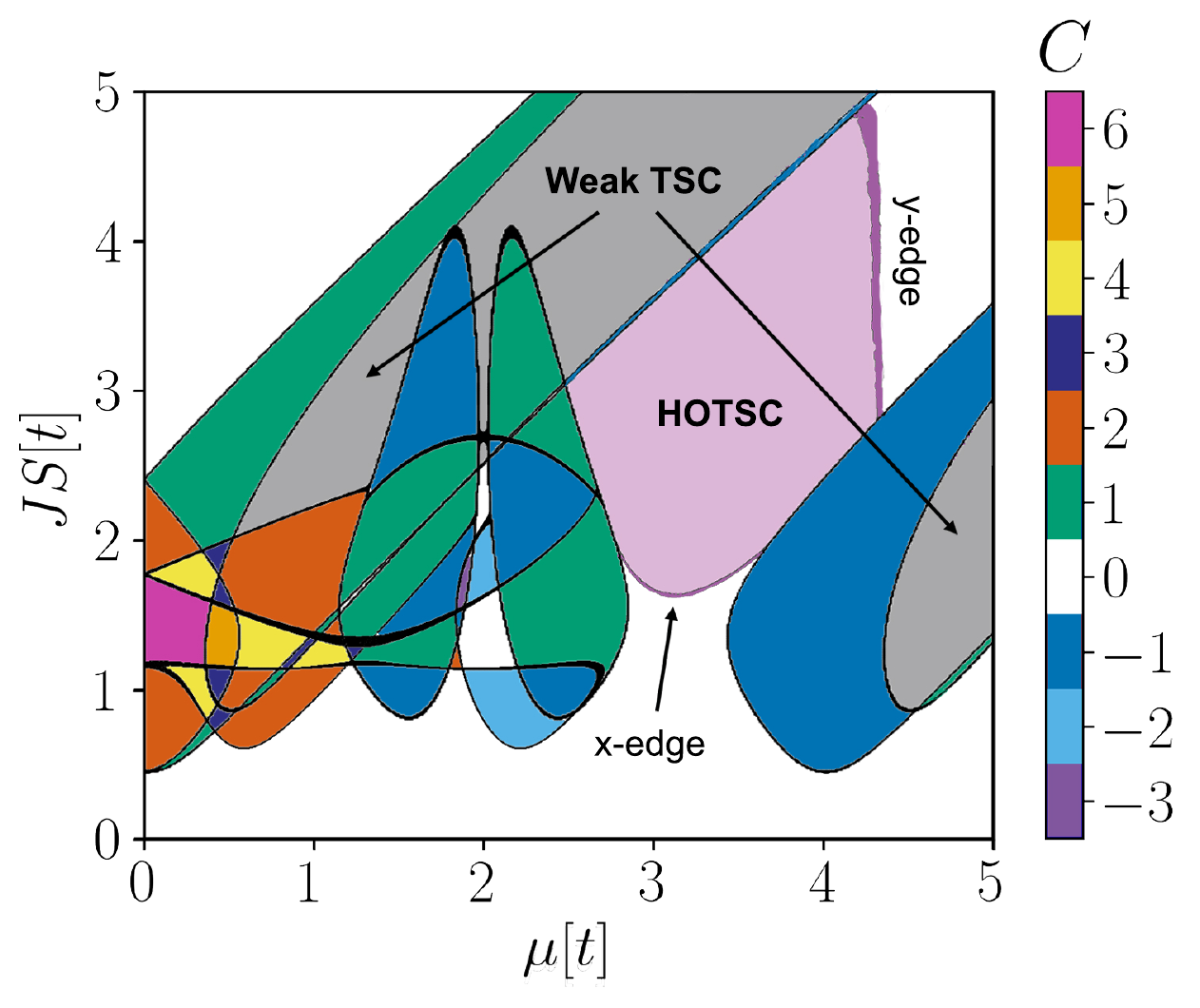}
        \caption{Topological phase diagram of an MSH system with $(J, J, 0)$ stacking as a function of $\mu$ and $JS$. Bulk and edge gap closings are indicated by black and purple lines, respectively.  Parameters are $\lambda=0.8t$ and $\Delta=0.3t$. The Chern number of strong TSC phases is indicated by the color bar.}
        \label{fig:App1}
\end{figure}

In Supplementary Fig.~\ref{fig:App1} we present the topological phase diagram of an MSH system with a $(J, J, 0)$ stacking, similar to the one presented in Fig.~\textbf{1d} of the main text, but with a smaller superconducting order parameter, $\Delta=0.3t$. In contrast to the topological phase diagram in Fig.~\textbf{1d} (where $\Delta=1.2t$), here the number of strong TSC phases has significantly increased, with Chern numbers ranging from $C=-3$ to $C=6$. The weak phases exhibited by this phase diagram match those of Fig.~\textbf{1d}, but the HOTSC phase is more similar to the phase diagram of the $(J, 0, J)$ stacking in Fig.~\textbf{1c}. This indicates that the effective coupling between adjacent adatom chains is weaker than the coupling between chains separated by a row of bare substrate. This is corroborated by the corresponding small-$\Delta$ phase diagram for the $(J, 0, J)$ stacking, which is similar to the $(J, J, 0)$ stacking phase diagram in Fig.~\textbf{1d} of the main text. We conclude that reducing the superconducting order parameter exchanges the roles of the $(J, J, 0)$ and $(J, 0, J)$ stacking, but otherwise does not change the qualitative nature of the topological phase diagrams discussed in the main text.\\

{\bf Supplementary Note 2: Topological Phases of an isolated adatom chain}\\
In the main text we argue that the $x$-edge gap closing separating the trivial and HOTSC phases arises from a topological phase transition in the un-paired adatom chains at the top and bottom edges of the 2D MSH island. Here we confirm this picture by calculating the phase diagram of a single adatom chain on a superconducting substrate, analytically and numerically, and by plotting the LDOS of a single adatom chain on a two-dimensional superconducting substrate.

Like the bulk Hamiltonian in Eq.~(1), a single chain of magnetic adatoms possesses particle-hole symmetry but not time-reversal symmetry, placing it in symmetry class D of the Altland-Zirnbauer classification~\cite{altlandzirnbauer, kitaev2009periodic, ryu2010topological}. In one dimension, such systems are classified by a $\mathbb{Z}_2$ topological invariant that counts the parity of the number of Majorana zero modes at system boundaries. We now calculate this invariant for a single adatom chain and show that it exhibits the expected topological phase transition.

\begin{figure*}
    \subfloat[][]{\label{fig:App2a}\includegraphics[width=0.30\textwidth]{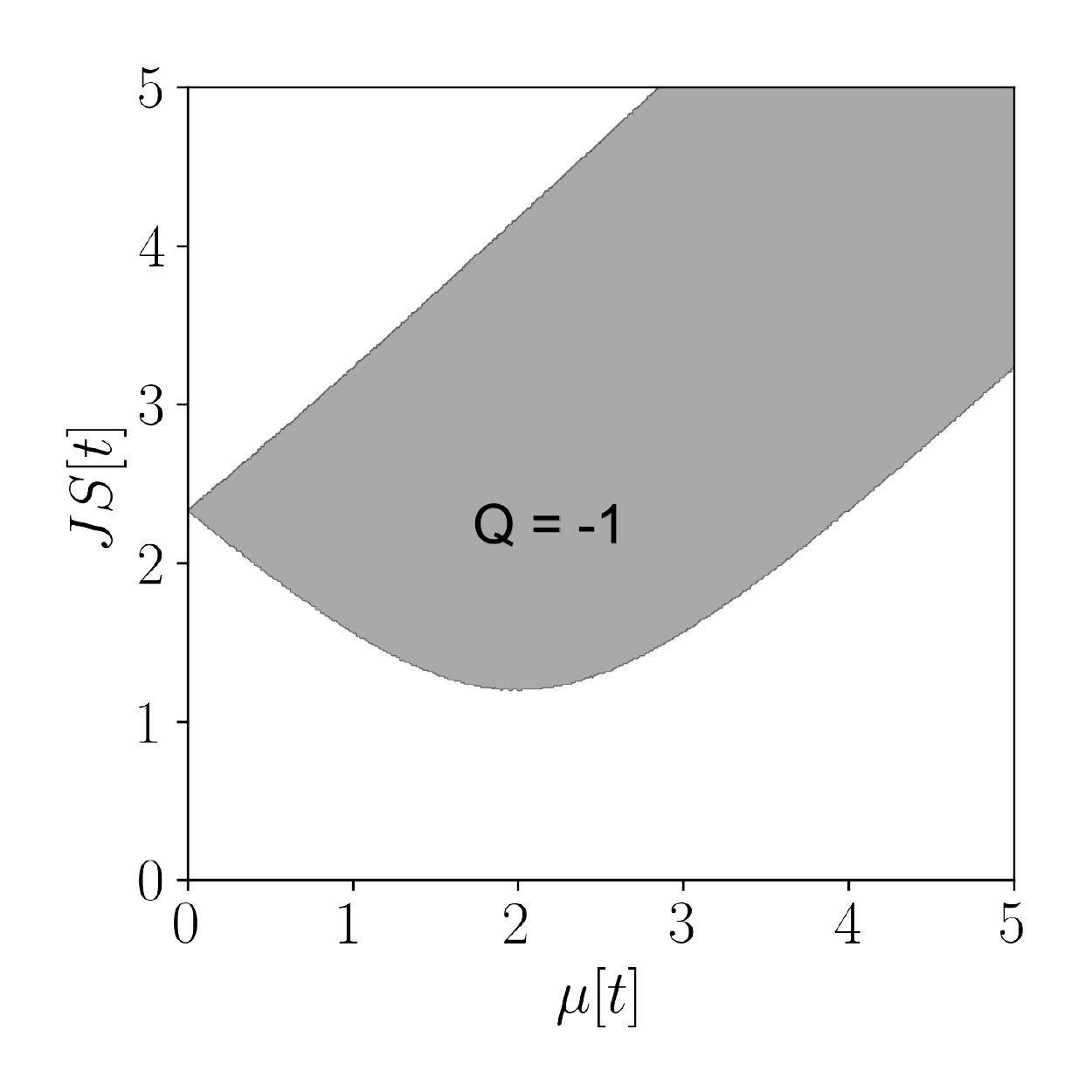}}
    \hfill
    \subfloat[][]{\label{fig:App2b}\includegraphics[width=0.30\textwidth]{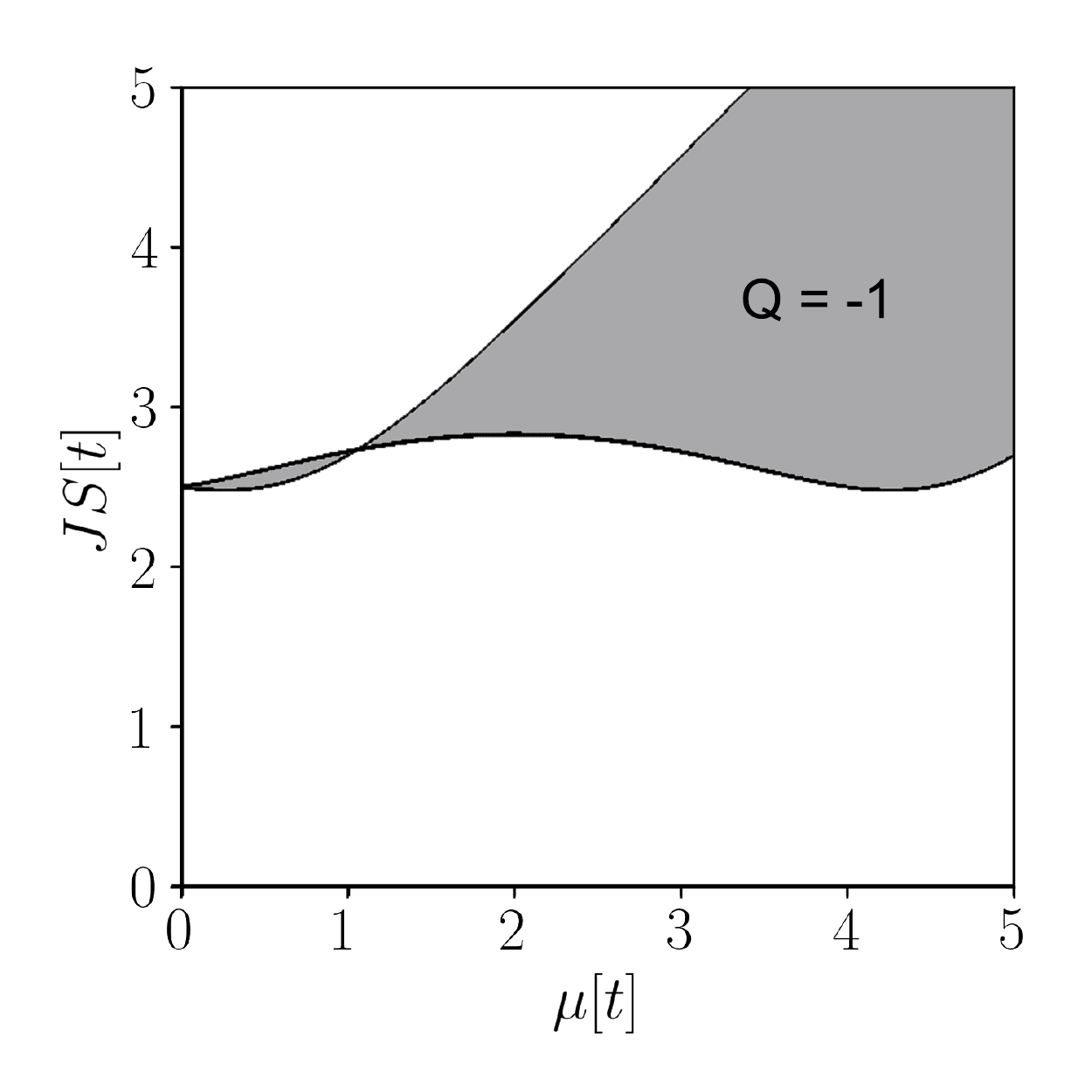}}
    \hfill
    \subfloat[][]{\label{fig:App2c}\includegraphics[width=0.30\textwidth]{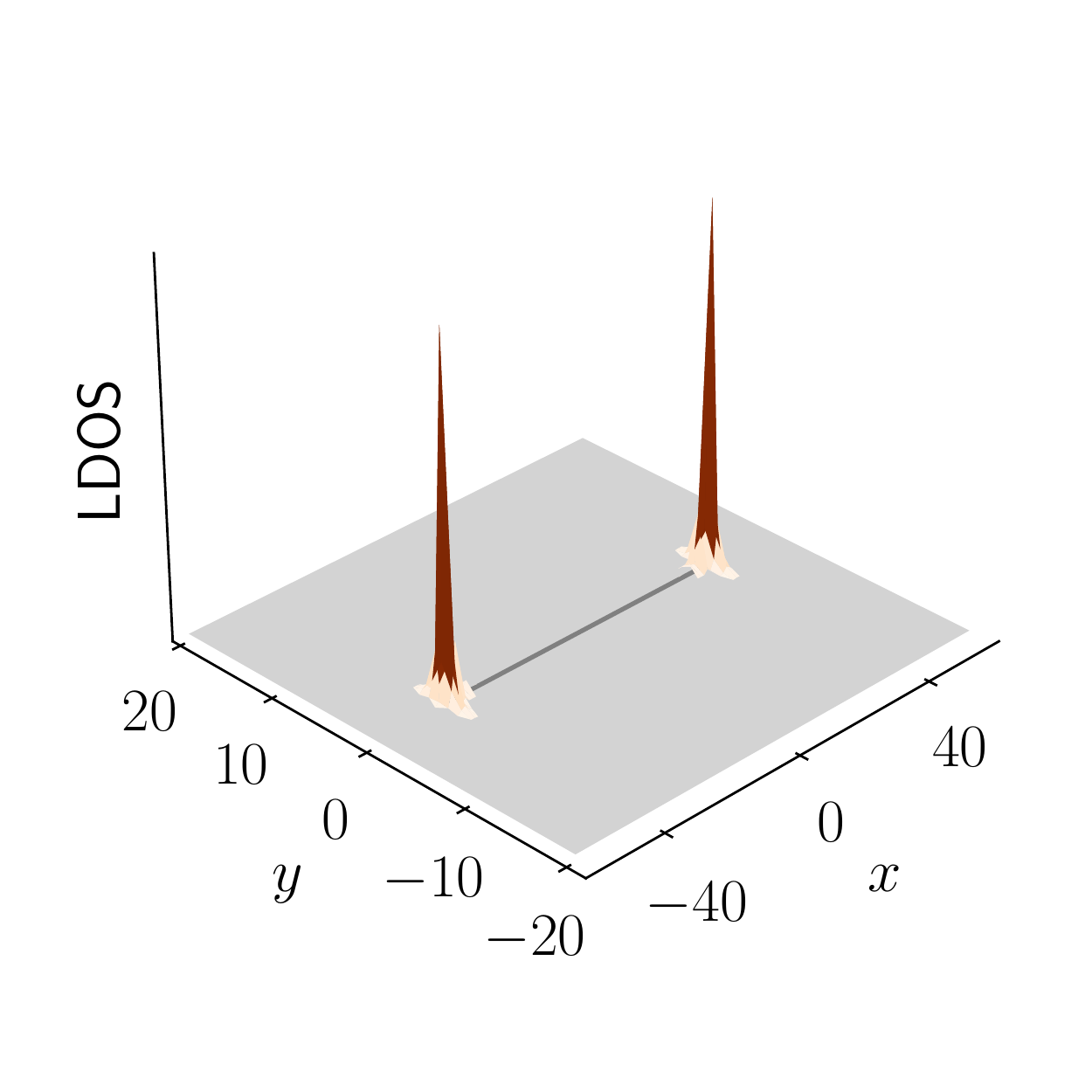}}
    \caption[]{
    {\bf a} Phase diagram of a single adatom chain on a one-dimensional superconducting substrate obtained from Eq.~\eqref{eq:single_chain_Q}. The gray region with $Q=-1$ represents the topological phase. {\bf b} Phase diagram of a single adatom chain on a two-dimensional superconducting substrate. The black lines indicate the bulk gap closings of the system calculated with periodic boundary conditions in the $x$-direction. The grey region denotes the topological phase in which the adatom chain hosts one Majorana zero mode at each of its ends.
    {\bf c} The zero-energy LDOS of a single adatom chain on a two-dimensional superconducting substrate in the topological phase (grey region of {\bf b}). The location of the adatoms is indicated by the filled dark gray circles, and the Majorana zero modes appear as sharp peaks at the ends of the chain. Parameters are $\lambda=0.8t$ and $\Delta=1.2t$, and $\mu=JS=4t$ for {\bf c}.
    }
    \label{fig:App2}
\end{figure*}

The Bloch Hamiltonian of a single adatom chain on a 1D superconducting substrate, as derived from Eq.~(1), is given by
\begin{equation}
    \begin{aligned}\label{eq:chain_ham}
        H(k) &= -(2t \cos{k} + \mu)\tau_z \sigma_0 + 2 \lambda \sin{k} \, \tau_z \sigma_y
        \\ &+ J \tau_z \sigma_z + \Delta \tau_y \sigma_y,
    \end{aligned}
\end{equation}
where $\tau_i$ and $\sigma_i$ are Pauli matrices acting on the particle-hole and spin degrees of freedom, respectively. Particle-hole symmetry ensures that \eqref{eq:chain_ham} can be placed in an antisymmetric form at time-reversal invariant momenta (TRIM) $k_0\in\{0,\pi\}$ via a unitary transformation,
\begin{equation}
    \label{eq:U_anti}
    \tilde{H}(k_0) = U  H(k_0) U^\dag, \quad U = \frac{1}{\sqrt{2}} \mqty(1 & 1 \\ i & -i) \otimes \sigma_0.
\end{equation}
The topological invariant of the chain is constructed from the Pfaffian of the antisymmetrized Hamiltonian at both TRIMs~\cite{Kitaev2001}:
\begin{equation}
    \label{eq:Q}
    \begin{aligned}
    Q &= \text{sgn}(\text{Pf}[i\tilde{H}(0)]\, \text{Pf}[i\tilde{H}(\pi)]).
    \end{aligned}
\end{equation}
The trivial phase corresponds to $Q=+1$ and the topological phase to $Q=-1$. For the Bloch Hamiltonian in \eqref{eq:chain_ham}, this invariant is given by
\begin{equation}
    \label{eq:single_chain_Q}
    \begin{aligned}
    Q &= \text{sgn}\left[(\Delta^2 - (JS)^2 + (\mu + 2t)^2)\right.\\
    &\left.\times  (\Delta^2 - (JS)^2 + (\mu - 2t)^2)\right].
    \end{aligned}
\end{equation}

In Supplementary Fig.~\ref{fig:App2a} we plot this invariant for the same parameters as the phase diagrams in Figs.~\textbf{1c}-\textbf{1d}. Moving along the $\mu=JS$ line to the top-right of the phase diagram, the trivial phase transitions to a topological phase that persists with finite extent around the $\mu=JS$ line to arbitrarily large $\mu$ and $JS$. This phase boundary corresponds to the $x$-edge gap closing in Figs.~\textbf{1c}-\textbf{1d}.

The mismatch between the precise location of the phase boundary in Supplementary Fig.~\ref{fig:App2a} and the $x$-edge gap closing line in Fig.~\textbf{1c} of the main text arises from the latter representing an MSH system on a 2D superconducting substrate. We confirm this by numerically calculating the phase diagram of a single adatom chain placed on a two-dimensional superconducting substrate in Supplementary Fig.~\ref{fig:App2b}, where the topological phase transition now shows better agreement with the $x$-edge gap closing in Figs.~\textbf{1c}. We also plot the zero-energy LDOS of this embedded adatom chain in the topological phase in Supplementary Fig.~\ref{fig:App2c}, further confirming that coupling to the two-dimensional bulk preserves the Majorana end states and only shifts the location of the phase boundary.\\

{\bf Supplementary Note 3: Topological phase diagram of a pair of  adjacent adatom chains}\\
Next, we study the phase diagram of an isolated pair of adjacent adatom chains on a superconducting substrate. The Hamiltonian for a pair of coupled adatom chains is also in symmetry class D and is therefore classified by the same invariant $Q$. Using the same procedure as for the single chain, we find that the invariant for a pair of coupled chains on a one-dimensional superconducting substrate is given by
\begin{equation}
    \label{eq:double_chain_Q}
    \begin{aligned}
    Q &= \text{sgn}[ (\Delta^2 - (JS)^2 + (\mu +t)^2 ) (\Delta^2 - (JS)^2 + (\mu +3t)^2 )
    \\ &\times (\Delta^2 - (JS)^2 + (\mu - t)^2 ) (\Delta^2 - (JS)^2 + (\mu - 3t)^2 ) ].
    \end{aligned}
\end{equation}
In Supplementary Fig.~\ref{fig:App3a}, we plot this invariant for
the same parameters as the phase diagrams in Figs. 1c-
1d. There are multiple regions in the phase diagram where the pair of coupled adatom chains effectively realizes a single topological Kitaev chain, hosting a single Majorana zero mode at each end. These regions correspond to the weak phases in Figs.~\textbf{1c}-\textbf{1d}, wherein the single Majorana modes at the end of each pair of adjacent chains hybridize to form a midgap Majorana band.

\begin{figure*}
    \subfloat[][]{\label{fig:App3a}\includegraphics[width=0.30\textwidth]{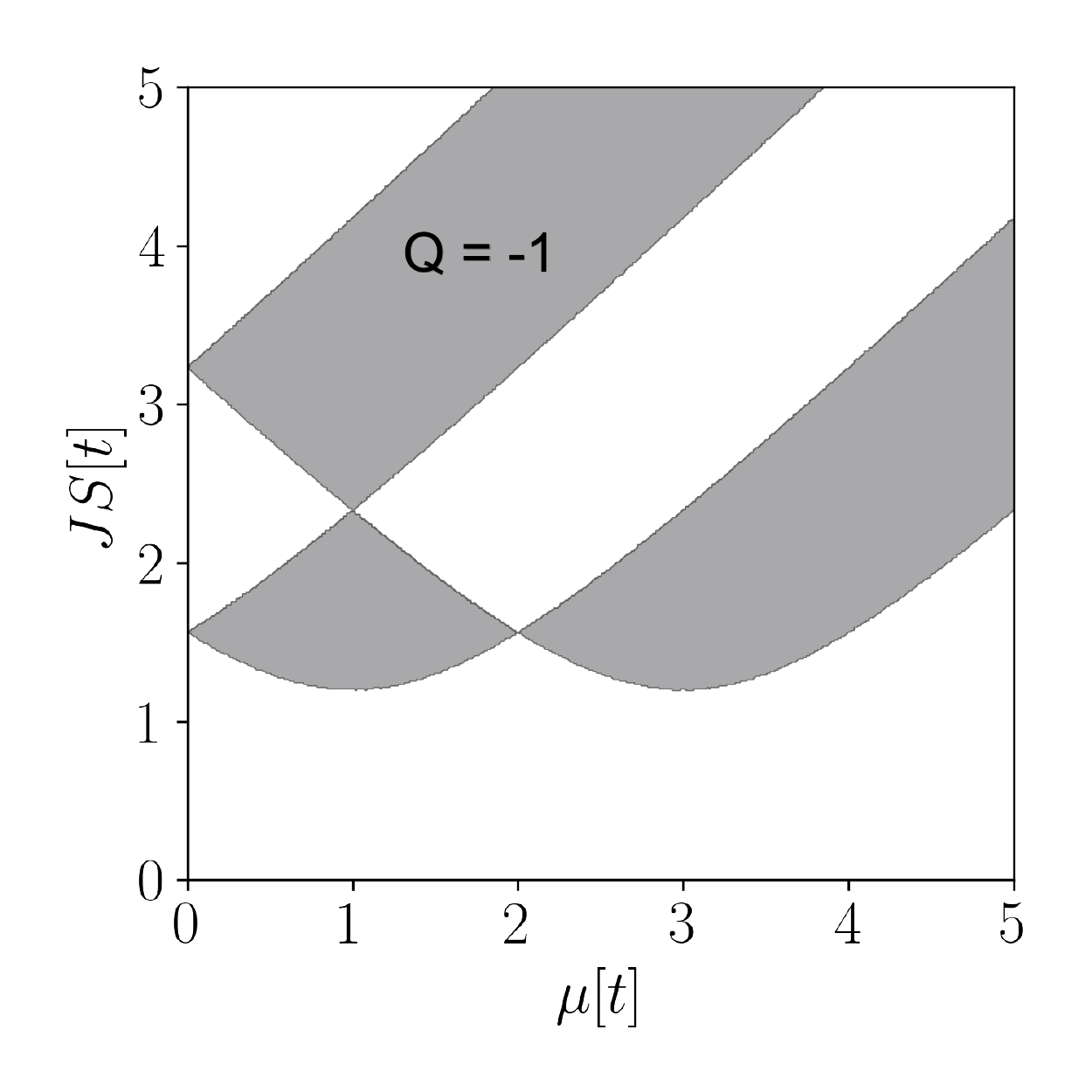}}
    \hfill
    \subfloat[][]{\label{fig:App3b}\includegraphics[width=0.30\textwidth]{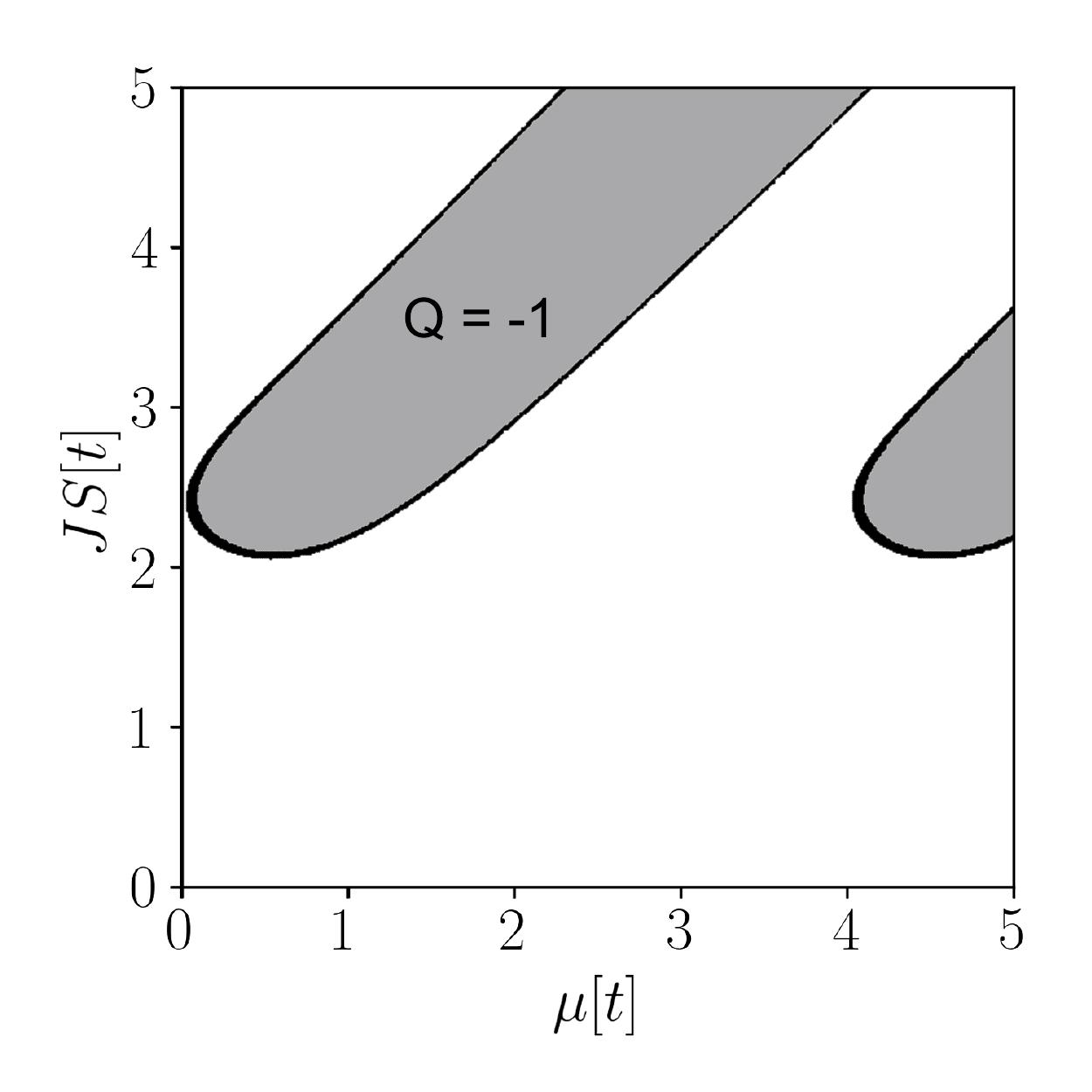}}
    \hfill
    \subfloat[][]{\label{fig:App3c}\includegraphics[width=0.30\textwidth]{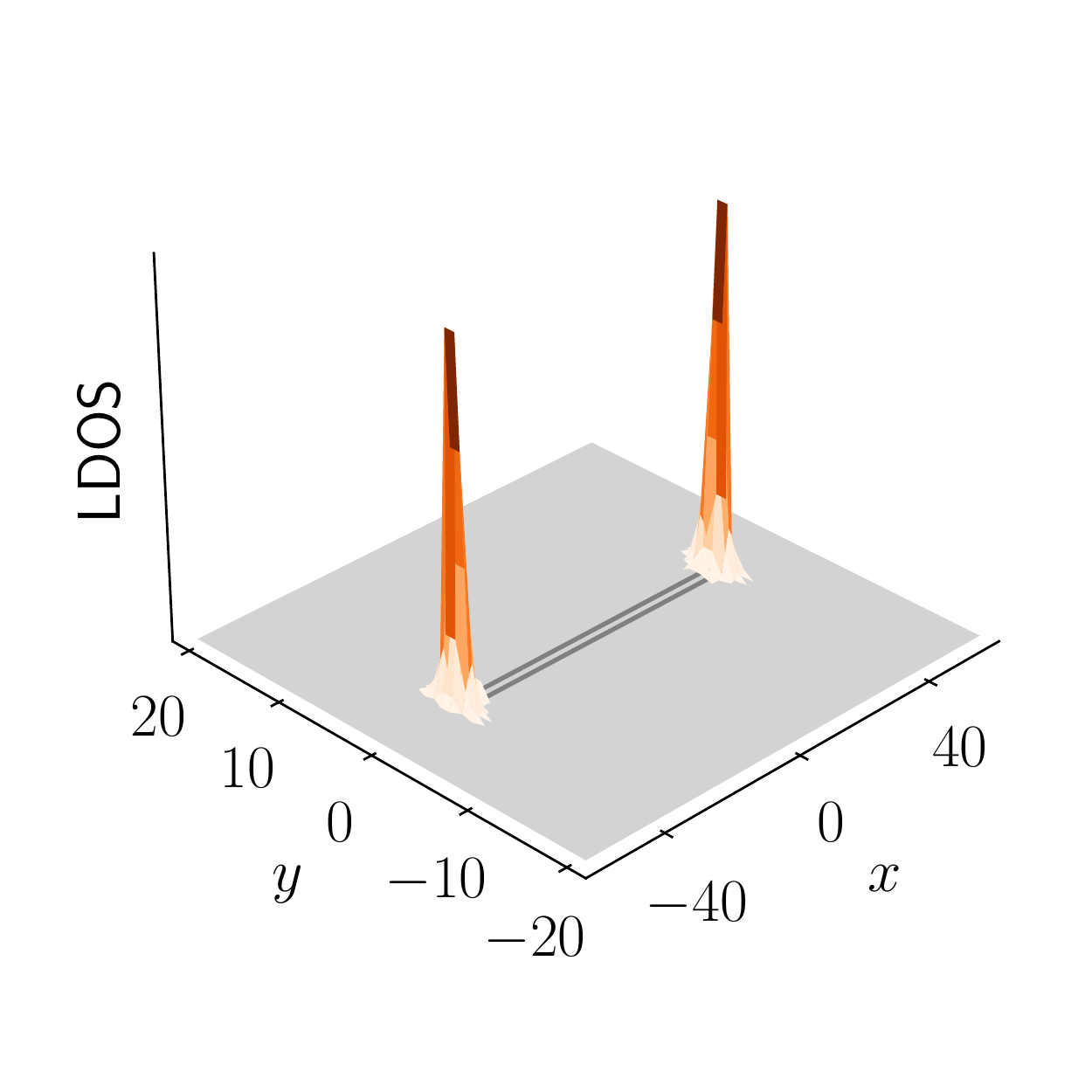}}

    \caption[]{
   {\bf a} Phase diagram of a pair of adjacent adatom chains on a one-dimensional superconducting substrate obtained from Eq.~\ref{eq:double_chain_Q}. The gray region denotes the topological phase with $Q=-1$. {\bf b} Phase diagram of a pair of adjacent adatom chains on a two-dimensional superconducting substrate. The black lines indicate the bulk gap closings of the system calculated with periodic boundary conditions in the $x$-direction. The grey regions denote the topological phase, in which the pair of adatom chains hosts a single Majorana zero mode at each of its ends. {\b c} The zero-energy LDOS of a pair of adjacent adatom chains in the topological phase on a two-dimensional superconducting substrate (grey region of {\bf b}). The location of the adatoms is indicated by the filled dark gray circles, and the Majorana zero modes appear as sharp peaks at the ends of the chain. Parameters are $\lambda=0.8t$ and $\Delta=1.2t$, and $\mu=2t$ and $JS=4t$ for {\bf c}.
    }
    \label{fig:App3}
\end{figure*}

As in the single chain case, the phase boundaries of the analytic result do not precisely match the phase boundaries of the weak phases in Figs.~\textbf{1c}-\textbf{1d}. In particular, the analytic result includes a topological region at small $\mu$ and $JS$ that does not correspond to a weak phase in Figs.~\textbf{1c}-\textbf{1d}. In Supplementary Fig.~\ref{fig:App3}{\bf b} we therefore plot the numerically obtained phase diagram of a pair of adjacent adatom chains on a two-dimensional substrate, finding that the topological phase at small $\mu$ and $JS$, which exhibits a very small gap, does not survive coupling to the two-dimensional substrate. We thus find that the weak TSC phase in Figs.~\textbf{1c}-\textbf{1d} lies within the $Q=-1$ phase of Fig.~\ref{fig:App3}{\bf b}. In Supplementary Fig.~\ref{fig:App3}{\bf c} we plot the zero-energy LDOS for a pair of adjacent adatom chains in the topological phase, confirming the presence of a single Majorana zero modes at each boundary of the pair.\\

{\bf Supplementary Note 4: Low-energy edge Hamiltonian near HOTSC transitions}\\
As a boundary-obstructed topological phase, the HOTSC is separated from the trivial phase only by edge gap closings. Passing through an edge gap closing corresponds to a change in the topology of the edge Hamiltonian. Near these gap closings, the edge Hamiltonians are well-approximated by Dirac Hamiltonians, and their topology is captured by the sign of the mass term. The edge transitions that bound the HOTSC phase must therefore involve a change in the sign of an edge mass.

Furthermore, a HOTSC phase can only arise when the $x$- and $y$-edges have different topology, and thus masses of opposite sign. In the rectangular MSH island, corners then act as domain walls between adjoining edges and trap Majorana zero modes. Here we numerically construct the low-energy edge Hamiltonian for the $(J,0,J)$ stacking to confirm this pattern of edge masses near the edge transitions.

We first study the $x$-edge transition in Fig.~\textbf{1d} and consider a cylinder geometry with periodic boundary conditions in $x$ direction and open boundary conditions in the $y$ direction with $N_y=60$ unit cells.
To construct the low-energy Hamiltonian, we diagonalize the semi-open Hamiltonian at an arbitrary point on the $x$-edge gap closing, $(\mu_0, J_0S) = (2.41, 2.36)t$. At this transition point, the Hamiltonian possesses four gapless bands that cross zero energy at $k_x=\pi$, two localized on each boundary. Projecting the Hamiltonian into the two zero-energy states on the top edge produces a gapless two-level effective edge Hamiltonian, $H_{\text{gapless}}(k_x)$.

We obtain the effective edge Hamiltonian at points away from the edge gap closing, $(\mu, JS) = (\mu_0 +\delta \mu, J_0S + \delta JS)$, by projecting the perturbed Hamiltonian into the two unperturbed zero-energy states at $(\mu_0, J_0S)$.
Expanding the edge Hamiltonian to first order in $(k_x-\pi)$, we obtain
\begin{equation}
    H_\text{edge} (k_x) =  H_\text{gapless}(k_x) + \sum_{j=1}^3 m_j(\delta \mu, \delta JS) \sigma_j,
\end{equation}
where $\sigma_j$ are the Pauli matrices. To compare the edge mass at different points in the phase diagram, we need a consistent definition of its sign. One valid choice is to take the sign of the edge mass to correspond to the sign of $m_2$. For simplicity, we can then use a unitary rotation to eliminate two of the three masses without changing this sign: ${U (m_1 \sigma_1 + m_2 \sigma_2 + m_3 \sigma_3) U^\dag = \tilde{m}_2 \sigma_2}$. Near the chosen critical point, we find
\begin{equation}
    \label{eq:y_edge_mass}
    \tilde{m}_{2, \text{x-edge}} \approx 0.07 \, \delta \mu + 0.26  \, \delta JS,
\end{equation}
which changes from negative in the lower-left trivial phase to positive in the HOTSC phase. Performing the same analysis for the $y$-edge, we find that the $y$-edge mass changes from negative in the HOTSC phase to positive in the upper-right trivial phase.  Therefore the two edge masses are of equal sign in the trivial phases but of opposite sign in the HOTSC phase, confirming the presence of corner domain walls and therefore Majorana zero modes.